\documentclass{jfm}
\usepackage{graphicx}
\usepackage{epstopdf, epsfig}
\usepackage{xcolor}
\usepackage{algorithm}
\usepackage{algorithmic}
\usepackage[hidelinks]{hyperref}
\usepackage{lineno}
\usepackage{amsmath}

\def\pmbf#1{\setbox0=\hbox{$#1$}%
        \kern-.025em\copy0\kern-\wd0
        \kern.05em\copy0\kern-\wd0
        \kern-.025em\raise.0433em\box0}
\newcount\ndots
\def\drawline#1#2{\raise 2.5pt\vbox{\hrule width #1pt height #2pt}}

\def\trian{\raise 1.25pt\hbox{$\scriptscriptstyle\triangle$}\nobreak\ }

\def\square{${\vcenter{\hrule height .4pt
        \hbox{\vrule width .4pt height 3pt \kern 3pt
        \vrule width .4pt}
        \hrule height .4pt}}$\nobreak\ }

\begin{document}

\shorttitle{Chiral particles in turbulence}
\shortauthor{Piumini, Assen, Lohse \& Verzicco}

\title{Particle chirality does not matter in the large-scale features of strong turbulence}
\date{}
\author{G. Piumini$^1\corresp{g.piumini@utwente.nl}$,
M.P.A. Assen$^1$,
D. Lohse$^{1,2}$ \&
R. Verzicco$^{1,3,4}\corresp{verzicco@uniroma2.it}$}

\affiliation{$^1$Physics of Fluids Group and Max Planck Center for Complex Fluid Dynamics, Department of Science and Technology, J.M. Burgers Center for Fluid Dynamics, and MESA+ Institute, University of Twente, P.O. Box 217, 7500AE Enschede, The Netherlands;\\
$^2$Max Planck Institute for Dynamics and Self-Organization, Am Fassberg 17, 37077 G{\"o}ttingen, Germany;\\
$^3$Dipartimento di Ingegneria Industriale, University of Rome ``Tor Vergata'', Via del Politecnico 1, 00133 Roma, Italy;\\
$^4$Gran Sasso Science Institute, Viale F. Crispi 7, 67100 L'Aquila, Italy.}

\maketitle
\begin{abstract}
We use three--dimensional direct numerical simulations of homogeneous isotropic turbulence in a cubic domain
to investigate the dynamics of heavy, chiral, finite--size inertial particles and their effects on the flow. 
Using an immersed--boundary method and a complex collision model, four--way coupled simulations have been
performed and the effects of particle--to--fluid density ratio, turbulence strength, and 
particle volume fraction have been analysed. We find that freely falling particles on the one hand add energy to the turbulent flow but, on the other hand, they also enhance the flow dissipation: depending on the combination of flow parameters,
the former or the latter mechanism prevails, thus yielding enhanced or weakened turbulence. Furthermore, particle 
chirality entails a preferential angular velocity which induces a net vorticity in the fluid phase. 
As turbulence strengthens, the energy introduced by the falling particles becomes less relevant and stronger 
velocity fluctuations alter the solid phase dynamics, making the effect of chirality irrelevant for the large-scale features of the flow. Moreover, comparing the time--history of collision events 
for chiral particles and spheres (at the same volume fraction) suggests that the former tend to entangle, in contrast
to the latter which rebound impulsively.
\end{abstract}

\section{Introduction}

\noindent Turbulent, inertial particle--laden flows are ubiquitous in natural and engineering environments 
\citep{geo}. Aerosols in clouds \citep{aero}, particle--driven gravity currents 
\citep{curr2005JFM}, sediments in oceans or rivers \citep{ocean}, sand storms in the atmosphere \citep{dust} 
as well as fluidisation of pipe or open--channel flows in industrial processes \citep{channel} are just a few examples 
among many. Owing to the 
high practical interest, investigation of these problems has increased in the last decade, especially thanks 
to the massive improvements achieved in terms of measurement and simulation performance. 
There are many aspects that make this topic an extraordinary challenge: the different properties of both 
particles and the carrier fluid, the multi-way coupling between continuous and dispersed phases and the 
relevant difference in spatial and time scales \citep{coletti, toschi}. The interaction between the two phases can affect the flow both at large and small scales depending on 
factors such as particle shape, size, and volume/mass fraction. Depending on their relative size to the 
viscous scale, heavy particles can attenuate or enhance turbulence \citep{modulation}, in contrast to
small, settling particles which always lead to turbulence damping \citep{dragred}. Moreover, particles of greater mass density than the surrounding fluid can detach from the flow, leading to preferential sampling, small-scale fractal clustering, and significant relative velocities \citep{gust}.

In recent years, studies have predominantly focused on particles with vanishing inertia (behaving as passive tracers) or highly symmetric shapes such as spheres or ellipsoids \citep{aero, dust, sozza}. However, investigations into anisotropic particles have gained traction only recently. Elastic fibers, for instance, induce similar turbulence damping as spheres but can also modify the energy scale-by-scale distribution \citep{olivieri,marchioli,giusto}. Inertial rods also possess the potential to influence turbulence-regeneration mechanisms, with rigid, inertia-less fibers generating additional stresses that dampen counter-rotating vortices in the wall region, leading to drag reduction. Inertial fibers may preferentially segregate into low-speed streaks, thereby delaying fluid flow \citep{soldati}.

Despite the abundance of literature in this field, most studies have focused on isotropic or highly symmetric particle shapes  \citep{coletti}, overlooking a significant portion of the parameter space concerning strongly asymmetric large particles in turbulence \citep{soldati}. The present paper is a first attempt aimed at filling this gap. Among several possibilities we have considered chiral particles which are distinct from their 
mirror image \citep{chiral}.
We have chosen chiral particles because they break spatial reflection symmetry and couple translational and 
rotational degrees of freedom. A particle forced to translate in a fluid will also rotate and vice versa; 
this feature is also present in nature and can be appreciated by watching the falling of maple seeds. Indeed, there are multiple examples of chiral objects in nature and technology. Most of flagella of bacteria and microorganisms are chiral but also propellers and impellers in mixers and industrial plants. \cite{chan} provides evidence that in nature there are more chiral objects than non chiral-ones. We wish to point out, however that our study is not focused on applications but rather on the effect of chiral objects on homogenous isotropic turbulence.  
\newline Questions we ask are: how do chiral particles behave in a turbulent flow? Are they able to modify some features
 of turbulence? What is the influence of parameters like solid--to--fluid density ratio, turbulence strength and
solid phase volume fraction?

The paper is organised as follows: in $\S$ \ref{sec:method} we outline the numerical model and the set-up of 
the simulations. The findings are discussed in $\S$ \ref{sec:results} and the complex interplay between particle 
and flow dynamics is illustrated by taking into account the effect of multiple parameters separately. 
Closing remarks and an outlook are given in $\S$ \ref{sec:conclusions}. 
The paper is completed with additional material in Appendix \ref{sec:appA}, \ref{appB}, and \ref{appC}.

\section{Problem and methods}
\label{sec:method}
\subsection{Setup}
\label{ssec:setup}
\begin{figure}
\centering
\includegraphics[width=\textwidth]{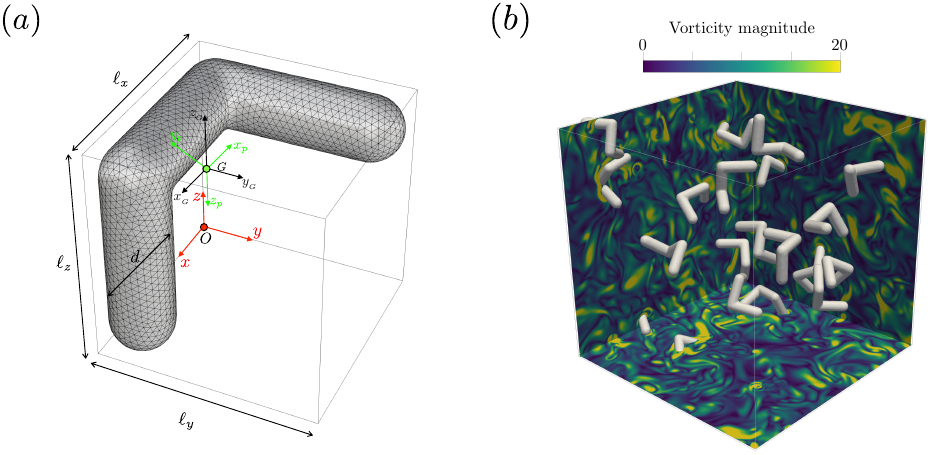}
\caption{(a) Representation of a chiral particle with global (red), body centroid (black) and principal inertia (green) reference frames.
  The green bullet (G) outside the particle is its center of mass. (b) Perspective view of the system with randomly distributed
 particles in homogeneous isotropic turbulence for $Re_\lambda=75$, volume fraction $\phi=1\%$ and density ratio 
$\rho_p/\rho_f=2$. Colours of the background are
nondimensional vorticity magnitude contours ranging from $0$ (blue) to $20$ (yellow).}
\label{fig:config}
\end{figure}

\noindent We consider a three--periodic cubic domain filled with a Newtonian fluid of density $\rho_f$ laden by $N_p$ identical 
chiral particles of density $\rho_p$. Each  particle is made of three perpendicular cylinders, arranged as in figure \ref{fig:config}a, 
with the ends rounded by spherical caps. Let $V_p$ be the volume of the particle and $D_{eq}$ the diameter of the equivalent sphere
(volumwise) which is assumed as unit length throughout the paper. The particle can be then enclosed in a cubic bounding-box of edges
$\ell_x$=$\ell_y$=$\ell_z$=$1.5 D_{eq}$ while the diameter of the cylindrical legs is $d \simeq 0.42 D_{eq}$. 
The particle is convex, as its centroid falls outside its volume, with coordinates $G$=$(0.75,0.42,1.08)$ in $D_{eq}$ units 
measured from the origin $O$. The asymmetric shape implies principal axes not aligned with any particle legs yielding
for their unit vectors $\widehat{\bf x}_p$=$(-0.774,0.447,0.447)$, $\widehat{\bf y}_p$=$(0.000,-0.707,0.707)$ and
$\widehat{\bf z}_p$=$(-0.633,-0.548,-0.548)$ and for the principal moments of inertia $I_{xp}=0.076$, $I_{yp}=0.226$ and $I_{zp}=0.248$ in
$\rho_p D_{eq}^5$ units.
It is worthwhile pointing out that the present particles are equivalent to a sphere of diameter $D_{eq}$ {\it only
volumewise}; in fact, the wetted  surface of the former is about $50\%$ bigger than the equivalent sphere while the
encumbrance of the boundingbox $\approx 3.4$ times that of a cube enclosing the sphere. These differences are
relevant for the particle/particle and particle/fluid interaction as will be discussed in the section of Results.

The flow domain is cubic with edges of length $L_x$=$L_y$=$L_z$=$10 D_{eq}$ in each direction and a constant gravitational acceleration $\bold{g}$ 
 anti--parallel to the  $z$ direction. Particles are initially distributed randomly, both in position and orientation, within the
computational domain, the
only constraint being that they cannot intersect spatially: a schematic of the system is shown in figure \ref{fig:config}b. 

After an initial transient, the system attains a statistical steady state over which data are collected
for the analysis; depending on the problem parameters, statistic convergence is achieved for different time
windows. We define the large--eddy--turnover time $T_e=u_{rms}^2/\varepsilon$ with the velocity fluctuations
$u_{rms}$ and the kinetic energy dissipation rate $\varepsilon$ of the turbulent flow  and the `observation
time' $T_o$ the integration time window (in $D_{eq}/U$ units, $U$ being the flow velocity scale): simulations
of homogenous isotropic turbulence (HIT) without particles have been run for $33 \leq T_o/T_e \leq 10^3$, while for the simulations 
with a single particle we had $510 \leq T_o/T_e \leq 860$. Finally the cases with $N_p=20$ particles have 
been integrated for $75 \leq T_o/T_e \leq 840$. For each simulation the duration of the integration has 
been decided by monitoring the quantities of interest and verifying that their values did not change appreciably
during the final part of the simulation.

Before concluding this section we wish to point out that in some studies dealing with settling particles 
in turbulence (see for example \cite{chouippe}) an elongated domain in the direction of gravity has been
employed in order to avoid possible correlation effects produced by the preferential velocity direction.
In this study we have preliminarily verified that, at least for the flow quantities considered in the
present analysis, the differences in the results obtained in a cubic and in an elongated domain are negligible;
selected results from these tests are given in Appendix \ref{appC}.

\subsection{Carrier phase}
\noindent The governing relations for the fluid phase are the incompressible Navier-Stokes equations and the continuity equation which in nondimensional form read 
\begin{equation}
\frac{\partial\bold{u}}{\partial t} + (\bold{u}\cdot\nabla)\bold{u} = 
-\nabla p  +\frac{1}{Re}\nabla^2 \bold{u} + \pmb{{\rm f}},\ \ \ \ \ 
\nabla \cdot \bold{u} = 0.
\label{eq:NS}
\end{equation}
Here $\bold{u}$ and $p$  are fluid velocity vector and kinematic pressure, respectively, while $Re=D_{eq}U/\nu$ is the Reynolds number 
based on the length $D_{eq}$ and the flow velocity $U$ scales. 
The last term in the first of equations (\ref{eq:NS}) is a vector containing two volume forces 
$\pmb{{\rm f}}=\pmb{f}_{_T}+\pmb{f}_{_P}$: $\pmb{f}_{_T}$ sustains
 the HIT flow while $\pmb{f}_{_P}$ is the immersed boundary method (IBM) forcing which accounts for the presence of solid particles and allows for the two--way coupling 
between solid and fluid phases. 
 
We numerically solve the governing equations (\ref{eq:NS}) with our in--house advanced finite difference code  
\href{http://www.afid.eu}{(``AFiD'')} which is extensively described and validated in \cite{AFiD,VamRod}. The main features are: 
spatial derivatives are approximated by conservative, second--order accurate finite--differences discretised on a staggered mesh 
which is uniform and homogeneous ($\Delta x\! =\!  \Delta y\! =\!  \Delta z\!=\!\Delta$). A combination of Crank--Nicolson and low--storage third--order 
Runge--Kutta schemes is used to integrate, respectively, the viscous terms implicitly and all other terms explicitly. Finally
pressure and momentum are strongly coupled through a fractional--step method as described in \cite{MOHANRAI199115,VERZICCO1996402}.

\subsection{Homogeneous Isotropic Turbulence forcing method}
\noindent The HIT forcing $\pmb{f}_{_T}$ is prescribed, similarly to \cite{chouippe}, adopting the method introduced by
\cite{ESWARAN1988257}. It allows to attain a statistically stationary velocity field by forcing large scale (small wavenumber) 
components. This is achieved through a random Uhlenbeck--Ornstein process via a vector $\pmb{\hat{b}}(\pmb{k},t)$ 
whose components are prescribed in time according to
\begin{equation}
\hat{b}_i(\pmb{k},t+\Delta t)=\hat{b}_i(\pmb{k},t)\Bigg( 1-\frac{\Delta t}{T_L}\Bigg)+e_i(\pmb{k},t)\Bigg( 2\sigma^2 \frac{\Delta t}{T_L}\Bigg)^{\frac{1}{2}}, \hspace{0.5 cm} i=1,2,3.
\end{equation}
Here $\pmb{k}=(l,m,n)$ is the wavenumber vector,
 $\Delta t$ the numerical integration time step and $e_i(\pmb{k},t)$ a complex random number drawn from a standardised Gaussian 
distribution, $T_L$ and $\sigma^2$ are time scale and variance of the random process.

The forcing term $\pmb{\hat{f}}_{_T}(\pmb{k},t)$, in Fourier space, is non--zero only in the band $|\pmb{k}| \leq k_f=2.3$ and the 
final forcing is  obtained by an orthogonal projection of $\pmb{\hat{b}}(\pmb{k},t)$
\begin{equation}
\label{eq:fourier}
\pmb{\hat{f}}_{_T}(\pmb{k},t)=\pmb{\hat{b}}(\pmb{k},t)-\pmb{k}(\pmb{k}\cdot\pmb{\hat{b}}(\pmb{k},t+\Delta t))/(\pmb{k}\cdot\pmb{k}),
\end{equation}
which is needed to ensure the divergence--free condition. 
More details about the HIT forcing can be found in \cite{ESWARAN1988257,chouippe}. 

Note that equation (\ref{eq:fourier}) defines the forcing in Fourier space while equations (\ref{eq:NS}) are 
solved in the physical space.
Therefore the former expression must be transformed into values to be prescribed at the 
Eulerian grid points.
To determine the field in physical space  $\pmb{f}_{_T}(\pmb{x},t)$, we compute the Fourier coefficients from equation (\ref{eq:fourier}),
with $\pmb{\hat{f}}_{lmn}(t)=\pmb{\hat{f}}(l,m,n,t)$,
from which the value at the discrete position $\pmb{x}_{i,j,k}=(i\Delta,j\Delta,k\Delta)$ is obtained by
\begin{equation}
\label{eq:space}
\pmb{{f}}_{_T}(\pmb{x}_{ijk},t)=\sum_{l=-N_f}^{N_f} \sum_{m=-N_f}^{N_f} \sum_{n=-N_f}^{N_f} \pmb{\hat{f}}_{lmn}(t) 
\exp(\j \pmb{k}\cdot \pmb{x}_{i,j,k}), ~~\forall i,j,k
\end{equation}
where  $N^3_f=27$ is the number of forced  wavenumbers and $\j$ the imaginary unit.

The HIT forcing and additional simulation parameters are reported in table \ref{tab:param} where it is also shown that the spatial
resolution is always smaller than the Kolmogorov scale. 

A validation of the flow solver is provided by computing
 the one--dimensional longitudinal energy spectrum $E_{11}(k_1)$ and the isotropy parameter 
\begin{equation}
\label{eq:iso}
I(k_1)=\frac{E_{11}(k_1)-k_1\partial E_{11}(k_1)/\partial k_1}{2E_{22}(k_1)},
\end{equation}
for a case at $Re_\lambda \approx 140$, 
the latter being the Reynolds number based on the Taylor micro--scale $\lambda=(15\nu u^2_{rms})^\frac{1}{2}$ and on the rms value
of the velocity fluctuations $u_{rms}$. 

\begin{figure}
\includegraphics[width=\textwidth]{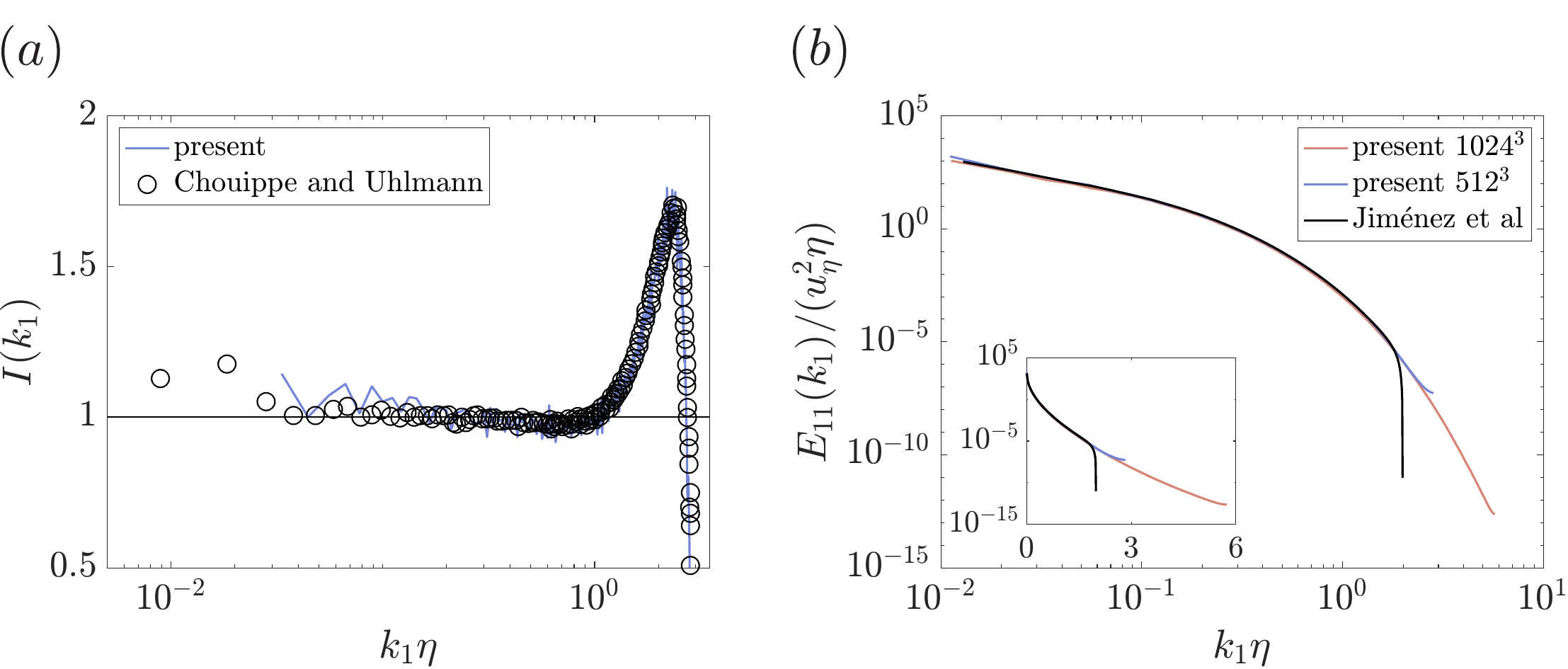}
\caption{Single phase HIT at  $Re_\lambda=140$: (a) isotropy parameter $I$, as defined in equation (\ref{eq:iso}), compared
 with \cite{chouippe}. (b) One--dimensional energy spectrum, compared with \cite{jimenez}.}
\label{fig:hitval}
\end{figure}

The comparison with the data by \cite{jimenez} and  \cite{chouippe} is given in
figure \ref{fig:hitval}, showing a very good agreement for both quantities.

As an aside, we note that results obtained with a resolution $512^3$ ($\eta/\Delta=0.92$) and 
$1024^3$ ($\eta/\Delta=1.83$) perfectly agree and, in both cases, the viscous range
shows an exponential decay, evidenced by the linear slope when
plotted in semi--log axes \citep{jimenez}. The reason for the sharp decay of the benchmark data 
in the viscous range is that they had been obtained using the spectral code
developed by  \cite{Rogallo}, which is very strict on isotropy. In fact, to approximate the 
flow as isotropically as possible, it applies every time step a mask in $k$--space,
such that the energy outside the sphere $|k\eta|=2$ is set to zero (J. Jim\'enez, {\it 
Personal Communication}).

\begin{table}
\begin{tabular*}{\linewidth}{@{\extracolsep{\fill}} ccccccccc } 
  RUN &  $T_L$             &$\sigma$     &      $Re_\lambda$  & $K$             & $\varepsilon$           & $\lambda/D_{eq}$ & $\eta/\Delta$ & $N^3$ \\ \hline \hline
 Re15 &$5.0\times 10^{-1}$ &$1.7\times 10^{-2}$& $13.0$       &$8.4\times 10^{-3}$&$2.7\times 10^{-4}$ & $1.74$           &  $6.30$       & $256^3$   \\
 Re30 &$8.5\times 10^{-1}$ &$6.2\times 10^{-3}$& $33.9$       &$1.1\times 10^{-1}$&$7.5\times 10^{-3}$ & $1.22$           &  $2.75$       & $256^3$   \\
 Re60 &$8.5\times 10^{-1}$ &$4.4\times 10^{-1}$& $66.7$       &$1.3\times 10^{0}~~$&$2.4\times 10^{-1}$ & $0.72$           &  $2.31$       & $512^3$   \\
 Re80 &$8.2\times 10^{-2}$ &$2.6\times 10^{0}~~$& $78.5$       &$2.2\times 10^{0}~~$&$5.3\times 10^{-1}$ & $0.64$           &  $1.90$       & $512^3$   \\
 Re140 &$8.5\times 10^{-2}$ &$8.3\times 10^{0}~~$& $134.0$       &$1.7\times 10^{+1}$&$1.0\times 10^{+1}$ & $0.40$           &  $1.83$       & $1024^3$   \\
\end{tabular*}
\caption{Run parameters of the single--phase simulations: 
$\lambda=(15\nu u^2_{rms}/\varepsilon)^{1/2}$ is the Taylor micro--scale while $Re_\lambda$ 
is the corresponding Reynolds number, $\eta=(\nu^3/\varepsilon)^{1/4}$ the Kolmogorov 
length scale, $N=N_x=N_y=N_z$ is the number of gridpoints of the Eulerian mesh in each direction, $K$ is the kinetic energy, and $T_L$ and $\sigma$ are the input parameters for the HIT forcing.}
\label{tab:param}
\end{table}

\subsection{Immersed boundary method}
\noindent As mentioned above, the term $\pmb{{\rm f}}$ on the right hand side of the first 
equation (\ref{eq:NS}) contains also 
the IBM part $\pmb{f}_{_P}$ which allows for two--way coupling between particles and fluid. Here we have  employed the method 
proposed by \cite{mls} in combination with the regularised Dirac delta function of \cite{roma}. 
The surface of each solid object is discretised by $N_l$ equilateral triangular elements (see figure \ref{fig:config}a) and Lagrangian markers are placed at their centroids. The IBM forcing term is computed on these Lagrangian markers, to satisfy velocity
boundary conditions, and then transferred back to the Eulerian mesh to impose the effect of the solid phase on the flow. In order to 
avoid unforced Eulerian nodes, which would result in surface `holes', the size of triangles edges must not exceed $0.7\Delta$ 
\citep{mls} thus implying that, for increasing $Re_\lambda$, not only the Eulerian mesh but also the Lagrangian triangulation
must be refined to maintain the correct phase coupling. Accordingly, the values of $N_l$ were $4912$, $10048$ or $21262$, depending
on the flow resolution.
 
It is worth mentioning that $N_p$ particles settling under their own weight impose on the fluid a force constantly aligned with gravity
which, in a tri--periodic domain, would produce  a continuous vertical acceleration of the flow and this would prevent the attainment
of a statistical steady state.  Following \cite{chouippe}, this is 
avoided by subtracting, at every time step,
the mean of the IBM terms applied by the particles on the flow, thus imposing  a force with zero
average in all directions: this implies that the net momentum of the system must be on average zero. Thus the downward motion
of the falling particles is compensated by an upward motion of the fluid such to give zero net vertical momentum
(this is the analogous of the time dependent pressure gradient in turbulent channel flows used to ensure a constant flow rate \citep{kimmoin}
in time). 
An analogous observation applies to angular momentum since free falling chiral particles rotate in a preferred direction 
and this would generate a net circulation. In a tri--periodic cube, however, the Kelvin-Stokes theorem imposes zero circulation
in all directions. Therefore the angular velocity of particles is compensated by fluid vorticity in the opposite direction
so to maintain the circulation zero at every instant.

These observations must be kept in mind when analysing the results since while the resulting system can be efficiently simulated
within homogeneous and isotropic conditions, an analogous laboratory experiment might not be immediately 
possible to set up.

\subsection{Particles}

\begin{figure}
\centering
\includegraphics[width=\textwidth]{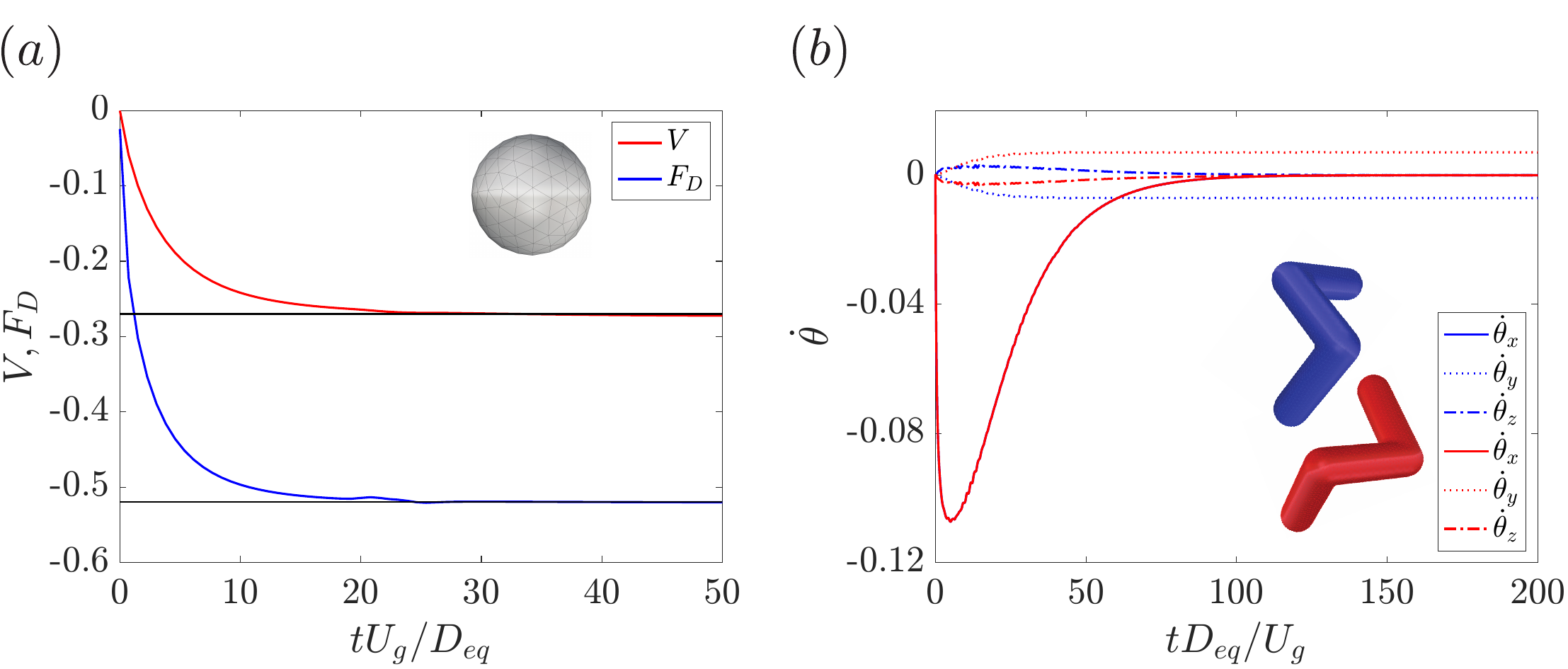}
\caption{Representative quantities for the Stokes dynamics of a free--falling sphere (a) and chiral particles (b) in a stagnant fluid.
(a) Vertical  velocity (red) and drag force (blue) are compared with the reference Stokes values (black). 
(b) Angular velocities in the body reference frame (green axes in figure \ref{fig:config}) for a left--handed (blue) and right--handed (red) chiral particles. Note that here, since the external turbulence forcing is absent,
the velocity scale $U$ is undefined; all results are therefore scaled using 
$U_g=(|\rho_p/\rho_f-1||\pmb{g}|D_{eq})^{1/2}$.}
\label{fig:coll}
\end{figure}

\noindent The dynamics of the dispersed solid phase is obtained by solving, for each particle,
 the six degrees of freedom of the Newton--Euler equations 
\begin{equation}
\label{eq:newton}
\hskip -0.1cm
\Bigg(
\rho_p V_p\frac{d\pmb{u}_c}{dt}\!=\!\oint_{A_p}\!\!\!\! \pmb{\tau}\cdot \pmb{n}dA+(\rho_p-\rho_f)V_p \pmb{g}+\pmb{F}_c \Bigg)^*\!\!,\ \ 
\Bigg(
\pmb{I}_p\frac{d\pmb{\omega}_c}{dt}+\pmb{\omega}_c\times(\pmb{I}_p\pmb{\omega}_c)\!=\!\oint_{A_p}\!\!\!\! \pmb{r}\times(\pmb{\tau}\cdot \pmb{n})dA+\pmb{T}_c\Bigg)^*\!\!,
\end{equation}
where the notation $(\cdot)^*$ implies a content in dimensional units.
Here $A_p$ is the particle wet surface, 
$\pmb{\tau}=-p\pmb{I}+\mu_f(\nabla\pmb{u}+\nabla\pmb{u}^T)$ 
the Cauchy stress tensor for a Newtonian fluid with $\pmb{I}$ the unit tensor, $\pmb{n}$ is the outward normal of the particle surface, 
 and $\pmb{I}_p$ the inertial tensor of the particle which, in the principal inertial reference frame (figure \ref{fig:config}a), 
has only the diagonal 
components.  $\pmb{F}_c$ and  $\pmb{T}_c$ represent, respectively, force and torque acting on 
the particle as result of collisions/physical contact with other particles. 

Following \cite{BREUGEM20124469} and relying on the third Newton's law of motion, we replace the integral terms on the 
right--hand--side of equations (\ref{eq:newton}) with the volume forcing of the fluid phase, obtaining
in nondimensional form: 
\begin{align}
\begin{split}
\frac{d\pmb{u}_c}{dt}\approx &\frac{\rho_f}{\rho_p}\frac{6}{\pi}\Bigg( -\sum_{i=1}^{N_l} \pmb{f}_i\Delta V_i
+\frac{d}{dt} \int_{V_p}\pmb{u}dV\Bigg) -\frac{\widehat{{\bf k}}}{Fr}
+\frac{\rho_f}{\rho_p}\frac{6}{\pi}\pmb{F}_c, \\
\pmb{I}_p\frac{d\pmb{\omega}_c}{dt}+\pmb{\omega}_c\times(\pmb{I}_p\pmb{\omega}_c)\approx&
\frac{\rho_f}{\rho_p}\frac{6}{\pi}\Bigg(- \sum_{i=1}^{N_l} \pmb{r}_i^n\times \pmb{f}_i\Delta V_i
+\frac{d}{dt}\int_{V_p}\pmb{r}\times \pmb{u}dV\Bigg)+\frac{\rho_f}{\rho_p}\frac{6}{\pi}\pmb{T}_c,
\label{eq:newton2}
\end{split}
\end{align} 
in which $\widehat{{\bf k}}$ is the vertical unit vector and
\begin{equation}
Fr= \frac{U^2\rho_p}{ D_{eq} |\pmb{g}|(\rho_p-\rho_f)}
\end{equation}
the Froude number.
In the above expression we note that surfaces are triangulated and each triangle is tagged
 by a Lagrangian marker at its centroid; thus the index $i$ indicates the Lagrangian point over 
the surface and  $\Delta V_i$ is the volume of the Eulerian cell intersected.

It is important to note that equation (\ref{eq:newton2}) clearly shows the particle dynamics to
depend on two nondimensional parameters, namely $\rho_p/\rho_f$ and $Fr$, which in principle
could be varied independently. 
Distinguishing between $\rho_p/\rho_f$ and $Fr$, however, might look
pleonastic since, in standard laboratory experiments within the same gravity
field, $\rho_p/\rho_f$ and $Fr$ are not independent and equation (\ref{eq:newton2})
would in fact depend only on a single parameter. Accordingly, for the vast majority of our simulations
we have fixed the magnitude of gravity ($|\pmb{g}|=2$) and varied only $\rho_p/\rho_f$ as independent parameter. This value of $|\pmb{g}|$ was initially selected since, for $\rho_p/\rho_f=2$, it yielded $Fr=1$.

However, centrifuge setups \citep{Jiang20} or
microgravity environments \citep{Futterer} could allow the independent variation of $\pmb{g}$, thus
making relevant studying the separate effect of $Fr$ and $\rho_p/\rho_f$.
In a few simulations, either of individual particles or of particle crowds, $\rho_p/\rho_f$ and $Fr$ have been 
varied independently, 
and a negligible effect of the former parameter compared to the latter has been observed;
Appendix \ref{sec:appA} reports some cases to confirm that indeed the results, for a fixed Froude number,
hardly depend on density ratio variations.

The above model has been validated by considering a free falling sphere in an initially quiescent viscous fluid. With a density ratio
$\rho_p/\rho_f=2$ ($Fr=1$) and a fluid viscosity such to yield $Re \approx 1$ in the stationary state, the Stokes solution is expected to
apply and this has been used for comparison. Domains up to $15\times 15\times 15$ in $D_{eq}$ units with a resolution of $288^3$
nodes have been employed and the results are given in figure \ref{fig:coll}a for the vertical velocity $V$ and the drag force $F_D$; 
the agreement between the present data and the analytical Stokes solution for analogous quantities looks excellent,
 once the stationary state is attained.
Similar calculations for the torque always yielded values within the round--off error as expected from the symmetry 
of the system geometry.
\begin{figure}
\centering
\includegraphics[width=\textwidth]{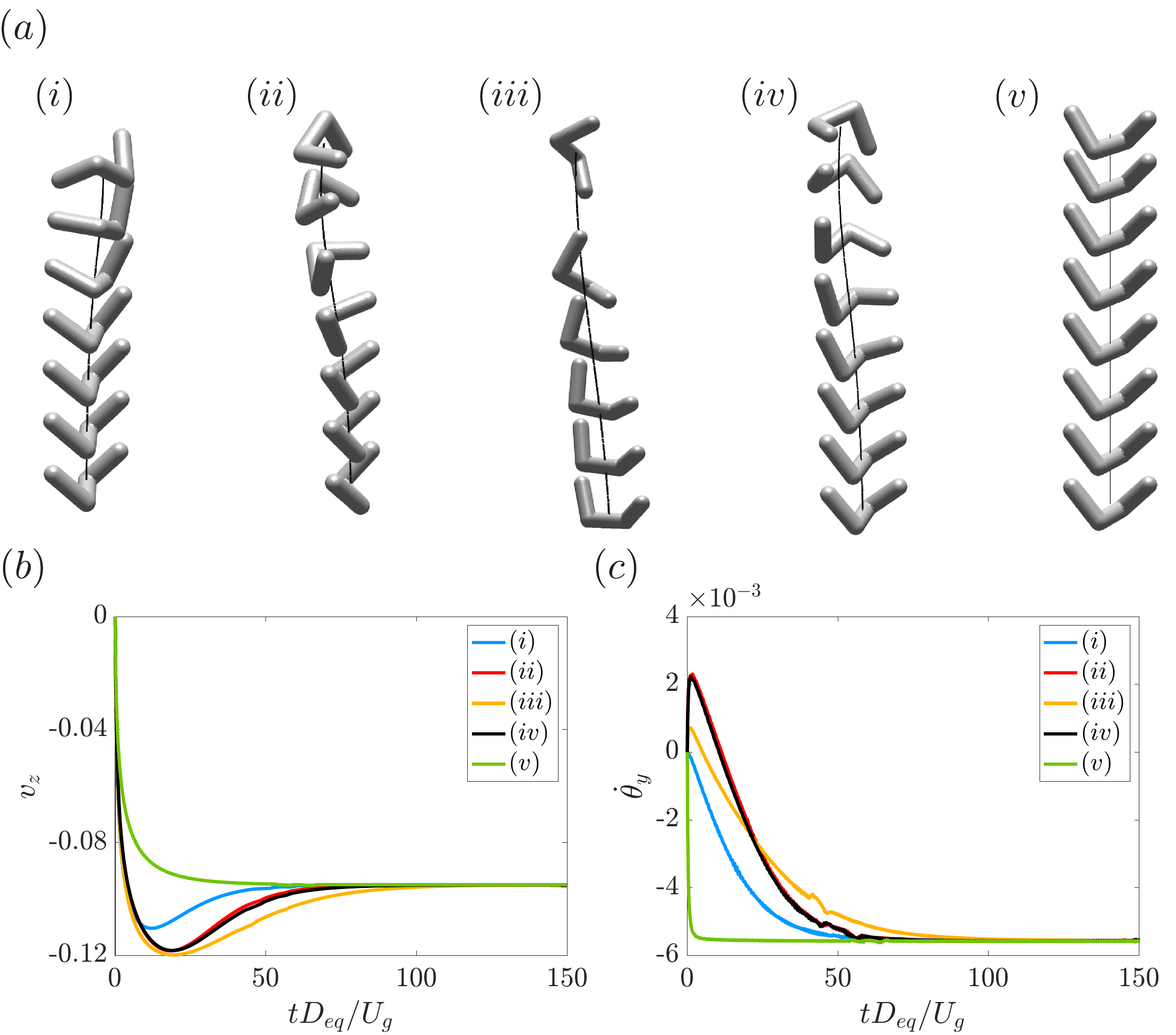}
\caption{Initial transient dynamics of a chiral particle falling in a stagnant fluid, for different initial orientations, at
$Re_p\approx 10$. The snapshots are taken at time lugs of $\Delta tD_{eq}/U_g=10$.
For the results of panels (b) and (c) the same scaling velocity as in figure \ref{fig:coll} has been used.
}
\label{fig:minco}
\end{figure}

Unfortunately, exact results in the low--$Re$ regime, as in figure \ref{fig:coll} a, are not available 
for chiral particles. Therefore we could only
run consistency checks, showing opposite steady rotation for left-- and right--handed particles.
More in detail, starting from any initial orientation, 
a free falling particle undergoes a transient dynamics during which rotations occur about all three axes until the mass--centroid attains 
the lowest possible position. Once this stable attitude is gained, only a constant rotation about the 
$y_p$ (free--falling) axis 
remains which has opposite sign for different chirality of the particle (figure \ref{fig:coll}b).
Additional simulations have revealed that the long--term dynamics is independent of the initial orientation which, instead,
affects only the duration of the transient dynamics (figure \ref{fig:minco}); this has important consequences on the particle 
interaction with turbulence, which will be discussed in the next Section. In order to further test this important finding, more simulations have been performed with decreased fluid viscosity so to obtain a particle Reynolds number up to $Re_p = 260$. In another series of tests the particle-to-fluid density ratio has been increased up to $\rho_p/\rho_f=10$ obtaining a particle Reynolds number up to $Re_p = 555$. For all cases, even if within strongly unsteady fluctuations, a similar dynamics as that described above has been observed thus confirming that the features are robust and mostly related to the specific particle geometry.

\begin{figure}
\centering
\hskip -1.0cm
\includegraphics[width=0.75\textwidth]{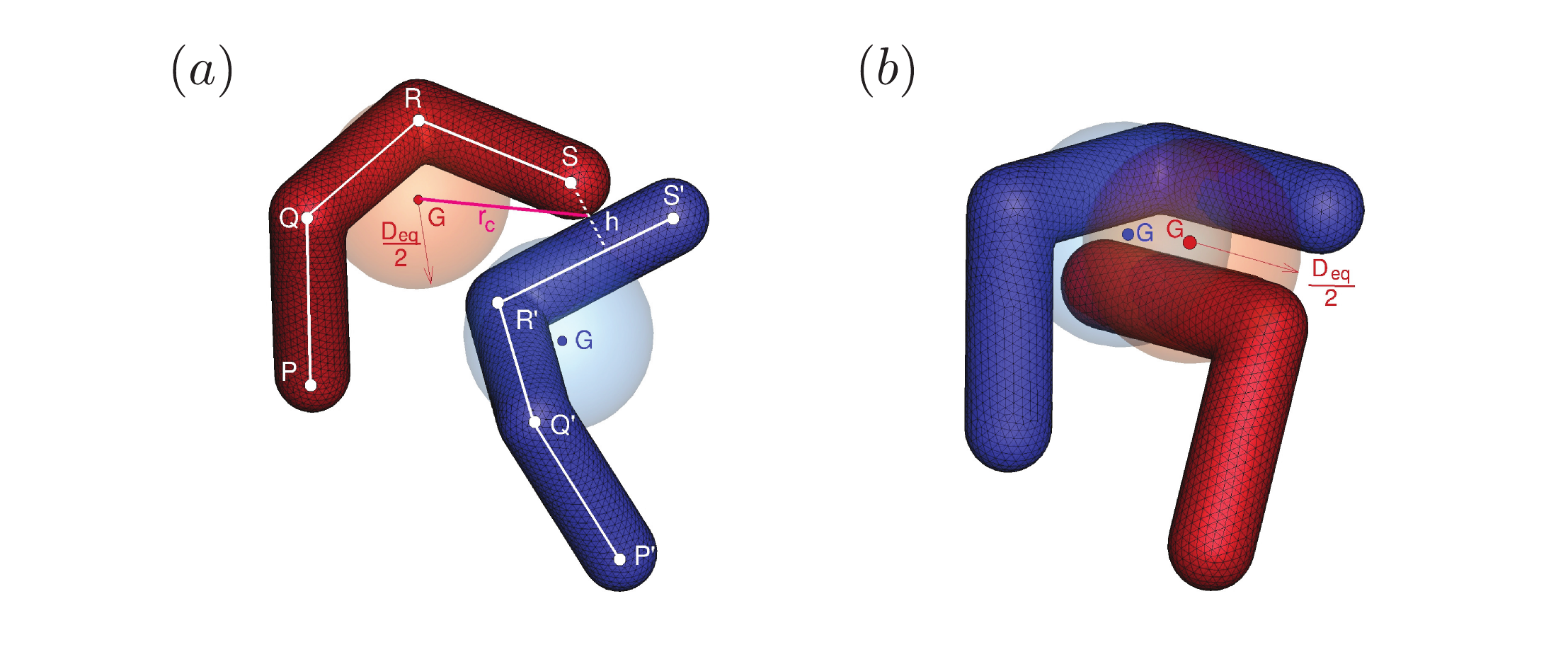}
\caption{Possible configurations for close particles; the volumwise equivalent spheres are represented by transparent solids:
(a) Particles with a contact point but with non--intersecting spheres.
(b) Particles not in contact with intersecting spheres. The white segments in panel (a) are the axes of the particles legs used to
compute proximity and contact.}
\label{fig:cont}
\end{figure}

An important aspect of the present problem is the evaluation of particle/particle collisions. 
In fact, we will see that, even if the volume fraction of the solid phase never exceeds $\phi=2\%$, the complex shape of particles involves
collisions with multiple contacts and even entanglement. Therefore the correct interaction dynamics is relevant for the
statistics of the whole system. 
Given the complex particle shape, determining interactions is not trivial as two bodies can have their centroids farther than $D_{eq}$ even
if there is contact (figure \ref{fig:cont}a)  or vice versa (figure \ref{fig:cont}b).

In fact, the collision model consists of two main ingredients: the proximity detection and the contact loads computation.
Concerning the former, each particle position and orientation is hard--coded through the coordinates of its  nodal points $P$, $Q$, $R$
and $S$ (figure \ref{fig:cont}a), the diameter $d$ of each leg (figure \ref{fig:config}a), and the radius $d/2$ of each distal 
spherical cap.
For any couple among the $N_p$ particles the centroids distance is computed and for those whose value is compatible
with a contact, the minimum distance among the segments $h$ is evaluated. Knowing the particle shape it is easy to compute the thickness
of the fluid gap between the two surfaces $d_c$ and, if it is smaller than two Eulerian grid cells, 
the contact loads are computed.
The reason for using $d_c\leq 2\Delta$ as threshold value is that, in IBM methods, the flow at 
the first external point is not obtained from the Navier--Stokes equations but from a model which enforces the boundary 
conditions; if the gap between two immersed boundaries
is thinner than $2\Delta$ only a single Eulerian point is available to satisfy two boundary conditions which is not
a well--posed problem.
Following \cite{BREUGEM20124469}, the short--range repulsive force, which the $j$--th particle exerts on the $i$--th,
 is prescribed according to 
$\pmb{F}_{c,ij}=-\alpha_c[(|\pmb{d}_{ij}|-d_c)/d_c]^2\pmb{d}_{ij}/|\pmb{d}_{ij}|$
with $\pmb{d}_{ij}$ the oriented minimum surface distance during the collision. 
$\alpha_c=10^{-4}$ as in \cite{BREUGEM20124469} and $\pmb{F}_{c,ji}=-\pmb{F}_{c,ij}$ as prescribed by the third Newton's law of motion.
Finally, the contact force produces also a contact moment which is immediately computed from
$\pmb{T}_{c,ij}= \pmb{r}_c \times \pmb{F}_{c,ij}$, with $\pmb{r}_c$ being the vector connecting the particle centroid with the
contact point (figure  \ref{fig:cont}a).

We refer to \cite{martin} for additional details and a thorough validation of the collision model.

\section{Results}
\label{sec:results}
\subsection{Turbulence intensity effect}
\label{sec:re}
\begin{figure}
\centering
\includegraphics[width=\textwidth]{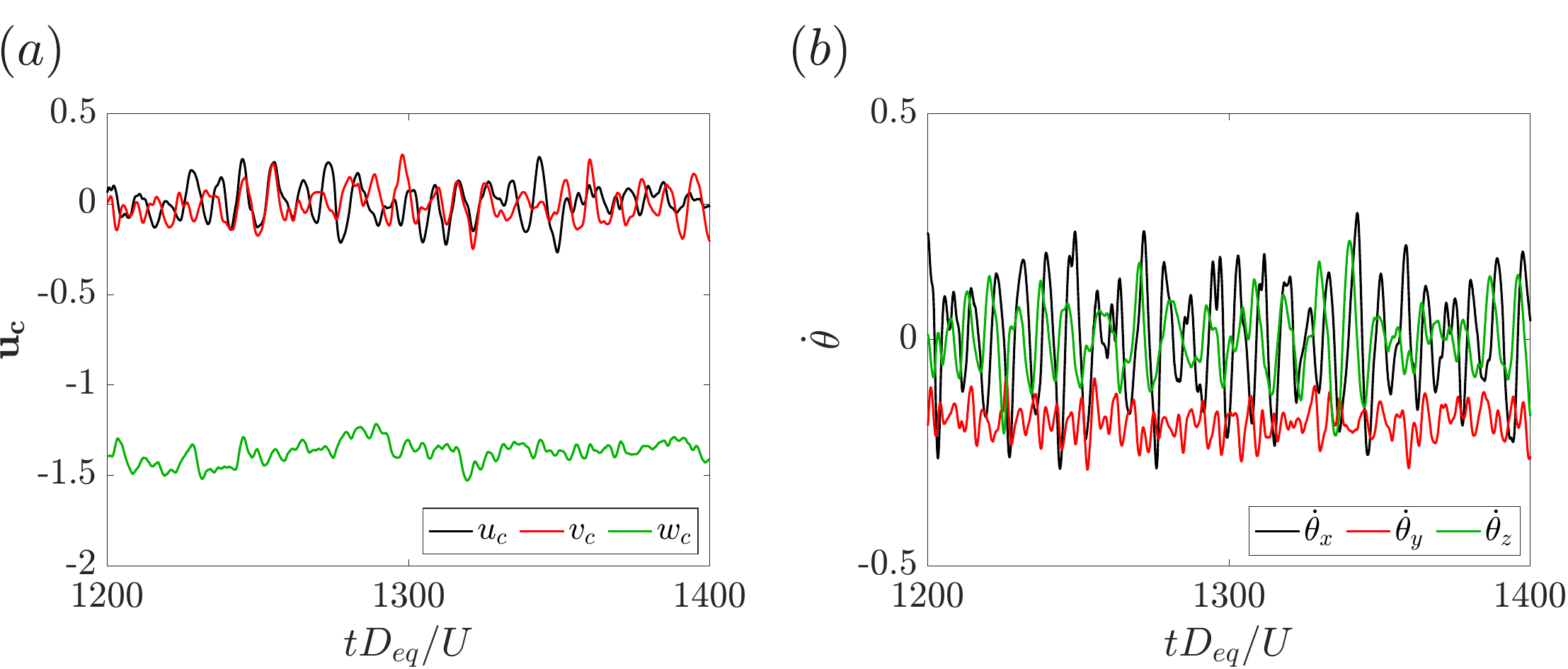}
\caption{Single chiral particle with $\rho_p/\rho_f=2$ ($Fr=1$) 
in a turbulent flow for the case Re15 of table \ref{tab:param}.
(a) Time evolution of the centre of mass velocity components. 
(b) Time evolution of angular velocity components expressed 
in the particle body frame (green axes in figure \ref{fig:config}). 
}
\label{fig:rot}
\end{figure}

\noindent In order to understand how chiral bodies interact with the surrounding flow, we preliminarily consider 
the case of a single particle (volume fraction $\phi=0.05\%$) in a turbulent environment of increasing
intensity.
We begin with a particle at $\rho_p/\rho_f=2$ ($Fr=1$) in weak turbulence, as for the case Re15 of table \ref{tab:param}.
figure \ref{fig:rot} shows also in this case a
net rotation about the $y_p$ body axis sustained by the vertical falling through the
 coupling of rotational and translational degrees of freedom.
\begin{figure}
\centering
\includegraphics[width=\textwidth]{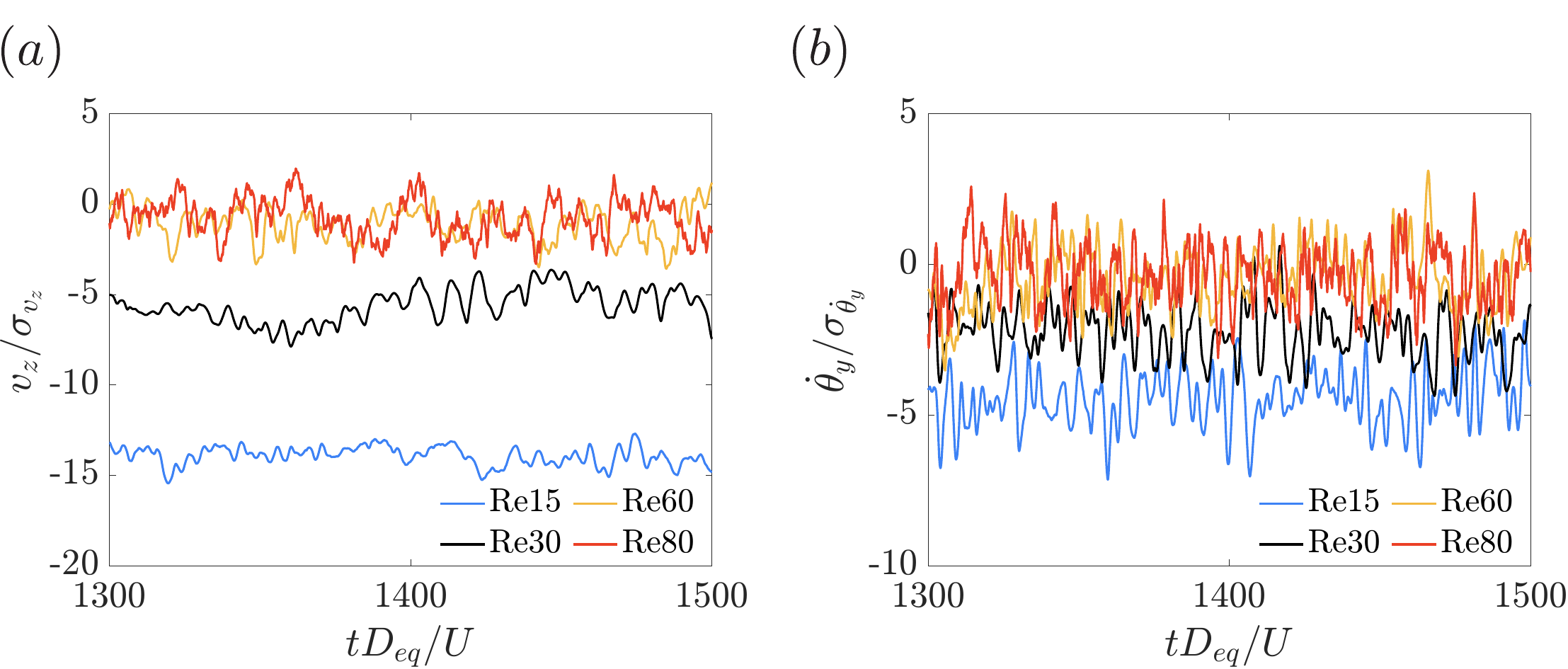}
\caption{Single chiral particle with $\rho_p/\rho_f=2$ ($Fr=1$) 
in a turbulent flow for different turbulence strengths (see table \ref{tab:param}).
(a) Time evolution of the centre of mass vertical falling velocity. 
(b) Time evolution of the angular velocity about
the $y_p$ particle body axis (green axes in figure \ref{fig:config}). 
}
\label{fig:rotn}
\end{figure}

Here the dynamics does not differ much from that of figure \ref{fig:coll}b, except
for the unsteadiness whose origin is twofold:
particle settling velocity and interaction with turbulence. In fact, $|\pmb{u}_c|$ 
is proportional to $U_g=(|\rho_p/\rho_f-1||\pmb{g}|D_{eq})^{1/2}$, in turn setting the particle 
Reynolds number $Re_p=U_g D_{eq}/\nu$ and, for high enough values, this yields
an unsteady falling even in a quiescent fluid  (for the present flow parameters it results 
$Re_p \approx 140$).
The interaction with turbulence, however, introduces further unsteadiness as the particle dynamics is coupled 
with velocity fluctuations whose strength is quantified by $Re_\lambda$. 
In this case it results $Re_\lambda\approx 14.7$ with a velocity ratio $u_{rms}/ |v_z| \approx 0.08$, suggesting that turbulence acts only as a perturbation
on the particle dynamics ($v_z = \pmb{u}_c \cdot \widehat{{\bf k}}$ is the falling particle velocity).

It is interesting to note that for the parameters of case Re15, single--phase
turbulence yielded $Re_\lambda \approx 13$ while the introduction of a particle
raises the value by more than $10\%$; this is due to the constant energy injection 
caused by the particle free fall (at the expenses of its potential energy) 
which adds to the external forcing and increases the turbulence strength.
Though the increase is small, the phenomenon underlines the two--way interaction
between fluid and solid phases whose dynamics is strongly connected.

As $Re_\lambda$ increases, the energy introduced by the particle
becomes less relevant to turbulence  while the stronger velocity fluctuations alter 
the solid body dynamics.
In fact, for the parameters of case Re80 the effective $Re_\lambda$ and $u_{rms}$
are indistinguishable from the single--phase values while particle mean vertical 
and angular velocities become negligible when compared to their fluctuations.

The results are summarised in figure \ref{fig:rotn}, which shows the mean vertical 
and  angular velocities, both normalised by their standard deviations. 
It is worth pointing out that the ratio $v_z/\sigma_{v_z}$ of figure \ref{fig:rotn}a
decreases with $Re_\lambda$ both because the standard deviation $\sigma_{v_z}$ increases and because
$|v_z|$ itself decreases. On the other hand, figure \ref{fig:rotn}b shows that  $|\dot \theta_y|$ remains approximately
constant within this range of $Re_\lambda$ although $\sigma_{\dot \theta_y}$ increases
and therefore their ratio decreases.

\begin{figure}
\centering
\includegraphics[width=\textwidth]{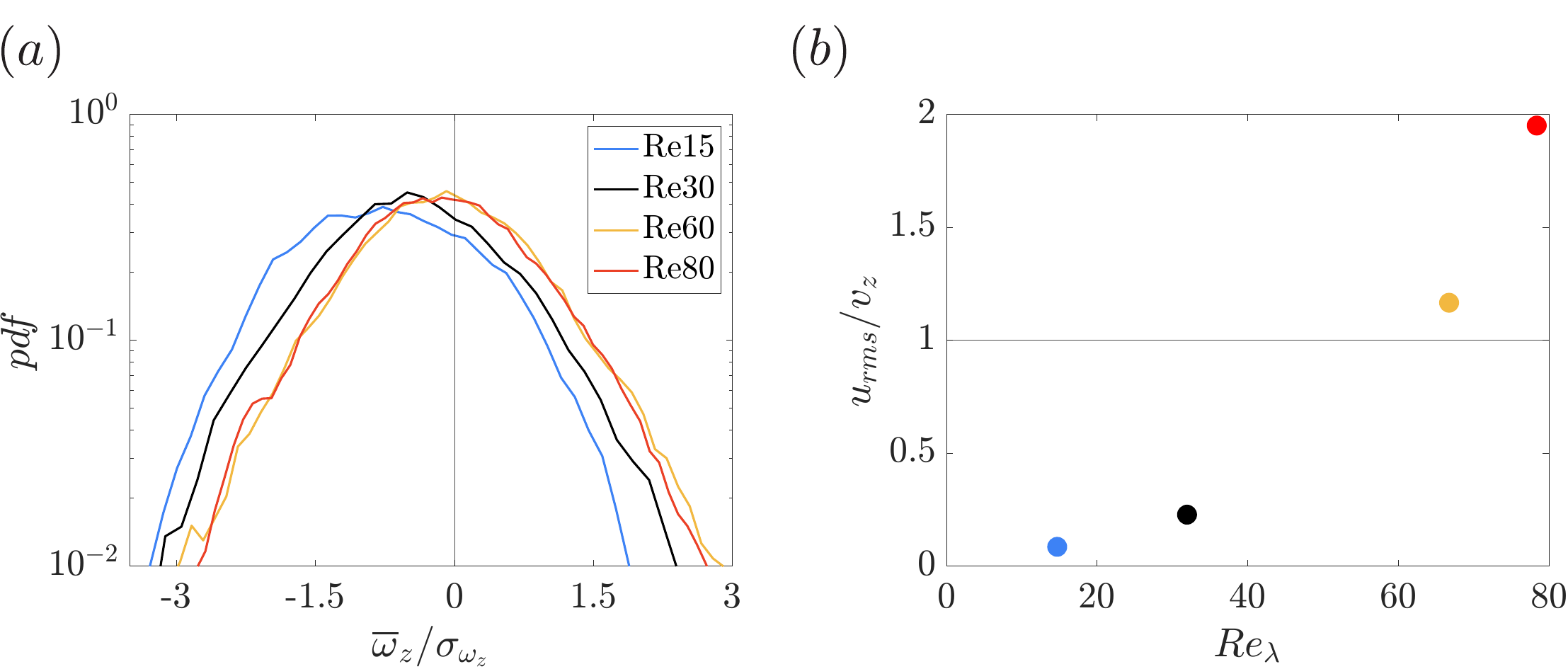}
\caption{Single chiral particle with $\rho_p/\rho_f=2$ ($Fr=1$) 
in a turbulent flow at different $Re_\lambda$.
(a) Probability density function of the fluid vertical vorticity component. A preferred vertical vorticity, due to the immersed chiral particles, is only obtained for weak turbulence.
(b) Ratio of turbulent fluctuation intensity and mean vertical particle velocity
as function of the effective flow $Re_\lambda$.
}
\label{fig:vorn}
\end{figure}

An important effect of particle chirality is its mean angular velocity which,
on account of the zero total circulation imposed by the domain boundary conditions, induces
an opposite mean vorticity in the flow field. This is indeed observed from the probability
density functions of the vertical fluid vorticity component (figure \ref{fig:vorn}a) which are
biased towards negative values and become more symmetric as $Re_\lambda$ increases and
particle rotation becomes less relevant.

These results are consistent with those of figure \ref{fig:rotn} suggesting that as
turbulence strengthens the particle tends to be passively advected by turbulent fluctuations
rather than altering turbulence itself through its helical free fall.

A partial confirmation of this conjecture comes from figure \ref{fig:vorn}b, which reports
the ratio between $u_{rms}$ and $|v_z|$: when this ratio is small particle dynamics 
dominates over turbulence which, therefore, is affected by particle chirality.
In contrast, large $Re_\lambda$ values results in $u_{rms}/|v_z| > 1$ and turbulent fluctuations 
dominate over particle forcing, yielding a more isotropic dynamics.

\subsection{Density ratio effect}
\label{sec:rho}

\noindent Given the role played by the relative magnitude of turbulence fluctuations ($u_{rms}$) and the
 particle falling velocity  $|v_z| \sim (|\rho_p/\rho_f-1||\pmb{g}|D_{eq})^{1/2}$,
it should be clear that the system is sensitive not only to the strength of turbulence forcing but also
to the density ratio $\rho_p/\rho_f$ (or $Fr$). In fact, the immediate
consequence of increasing $\rho_p/\rho_f$ is to enhance the falling velocity and therefore the
mean angular velocity. However, translation and rotation in a chiral particle are not rigidly 
coupled (as a nut and bolt assembly), as both depend on hydrodynamic loads. Therefore they
vary by different amounts. Moreover, the potential
energy transfer of the falling particle grows with $\rho_p/\rho_f$, thus further increasing
the turbulence level which reacts back on the particle dynamics.
Finally, $\rho_p/\rho_f$ changes particle inertia to translation and rotation, in turn 
altering the trajectory sensitivity to turbulent fluctuations.
\begin{figure}
\centering
\includegraphics[width=\textwidth]{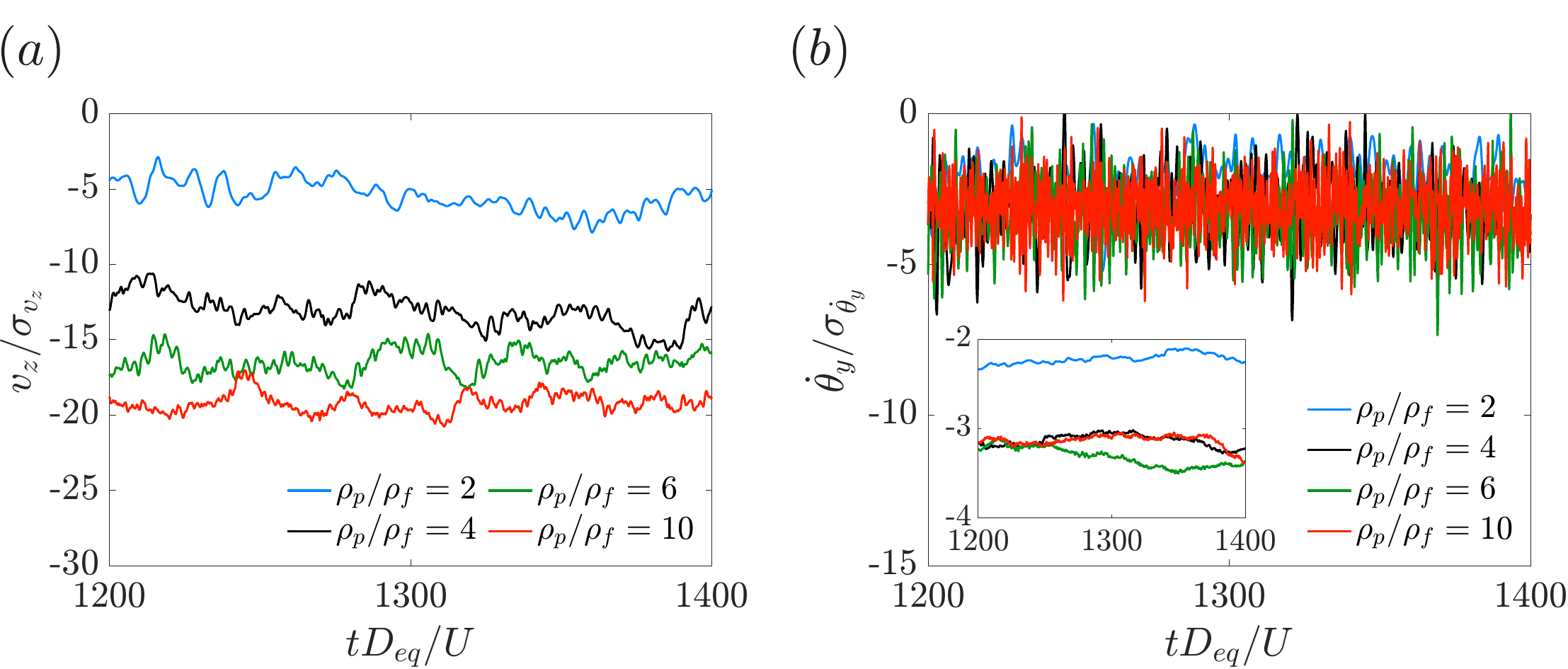}
\caption{Single chiral particle with $\rho_p/\rho_f$ from $2$ to $10$ 
($Fr$ from $1$ to $5/9$) in a turbulent flow 
with the forcing of Re15 case of 
 table \ref{tab:param}:
(a) Time evolution of the centre of mass velocity components. 
(b) Time evolution of angular velocity components expressed 
in the particle body frame. In the inset the same curves are
plotted with a running average of $20D_{eq}/U$ time units
to evidence their mean values.
}
\label{fig:drv}
\end{figure}

Examples of particle dynamics are given in figure \ref{fig:drv}  where results at $\rho_p/\rho_f$ from $2$ to  $10$ 
($Fr \in [1,5/9]$) are compared for
a turbulent forcing as the Re15 case of  table \ref{tab:param}: since for these flows the interplay between mean values and
standard deviations is more complex, we complement figure \ref{fig:drv} with table \ref{tab:drv}
where additional data are provided. 
A clear trend can be observed for the mean vales of  $v_z$ and $ \dot \theta_y $   
which both increase in magnitude with the ratio $\rho_p/\rho_f$. Also the effect of energy transfer, from particle to
turbulence, is clearly evidenced from the growth of $Re_\lambda$ and $u_{rms}$. A similar monotonic increase is
found for the mean vertical component of the fluid vorticity which is a direct consequence of the chirality
induced angular velocity of the particle.

\begin{figure}
\centering
\includegraphics[width=\textwidth]{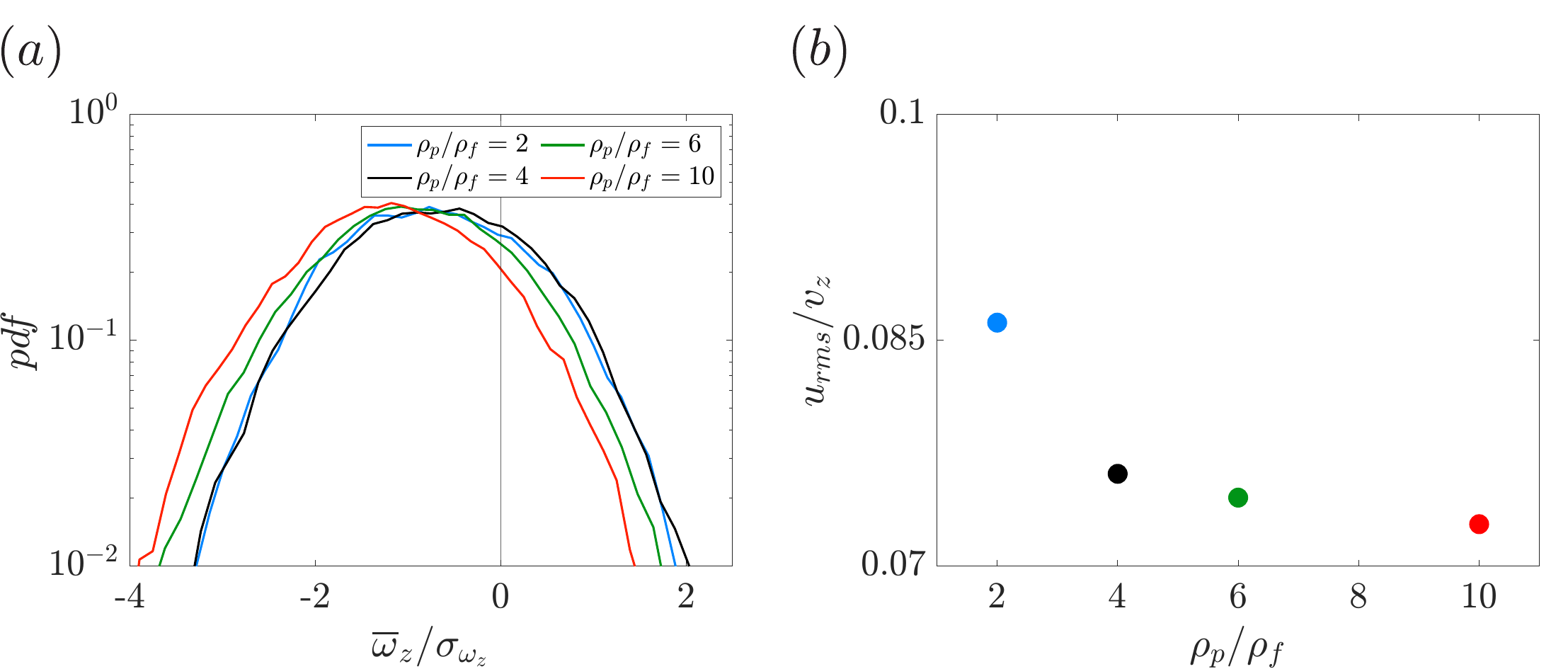}
\caption{
Single chiral particle with $\rho_p/\rho_f$ from $2$ to $10$ 
($Fr$ from $1$ to $5/9$) in a turbulent flow
with the forcing of Re15 case of
 table \ref{tab:param}:
(a) Probability density function of the fluid vertical vorticity component.
(b) Ratio of turbulent fluctuation intensity and mean vertical particle velocity 
as function of $\rho_p/\rho_f$.
}
\label{fig:vordrv}
\end{figure}

A closer look at the data, however, reveals that the fivefold  increase of $\rho_p/\rho_f$ produces growths
of   $v_z$   of $\approx3.6$ and    $\dot \theta_y$   of $\approx 2.7$, thus confirming the loose coupling
of particle rotation and translation.
Concerning the effect on turbulence,   $\overline{\omega}_z/\sigma_{\omega_z}$   grows by $\approx 1.6$. This is
confirmed by the histograms of figure \ref{fig:vordrv}a which drift in the direction of  negative 
vorticity consistently with the monotonic increase of  $ \dot \theta_y$.
On the other hand  figure \ref{fig:vordrv}b shows how, despite the growth of $Re_\lambda$, the ratio  $u_{rms}/ |v_z|$  decreases with
$\rho_p/\rho_f$,  consistently with the observation  that particles of larger density are less sensitive to turbulent
fluctuations and their chirality makes them behave as impellers which impart rotation to the
surrounding fluid.

\begin{table}
\begin{tabular*}{\linewidth}{@{\extracolsep{\fill}} cccccc}
  $\rho_p/\rho_f$ &  $\langle v_z\rangle$ & $\langle \dot \theta_y \rangle$ & $Re_\lambda$  & $u_{rms}$ & $ \overline{ \omega}_z/\sigma_{\omega_z}$   \\ \hline \hline
      $2$         &  $-1.41$ &        $-0.17$     & $14.7$        &    $0.12$ &      $-0.74$                    \\
      $4$         &  $-2.80$ &        $-0.29$     & $20.4$        &    $0.21$ &      $-0.67$                    \\
      $6$         &  $-3.73$ &        $-0.36$     & $23.4$        &    $0.28$ &      $-0.93$                    \\
      $10$        &  $-5.16$ &        $-0.46$     & $27.4$        &    $0.38$ &      $-1.17$                    \\
\end{tabular*}
\caption{Flow parameters of single chiral particle simulations with variable $\rho_p/\rho_f$ and
turbulent forcing as case Re15 of table \ref{tab:param}. }
\label{tab:drv}
\end{table}

\begin{figure}
\centering
\includegraphics[width=\textwidth]{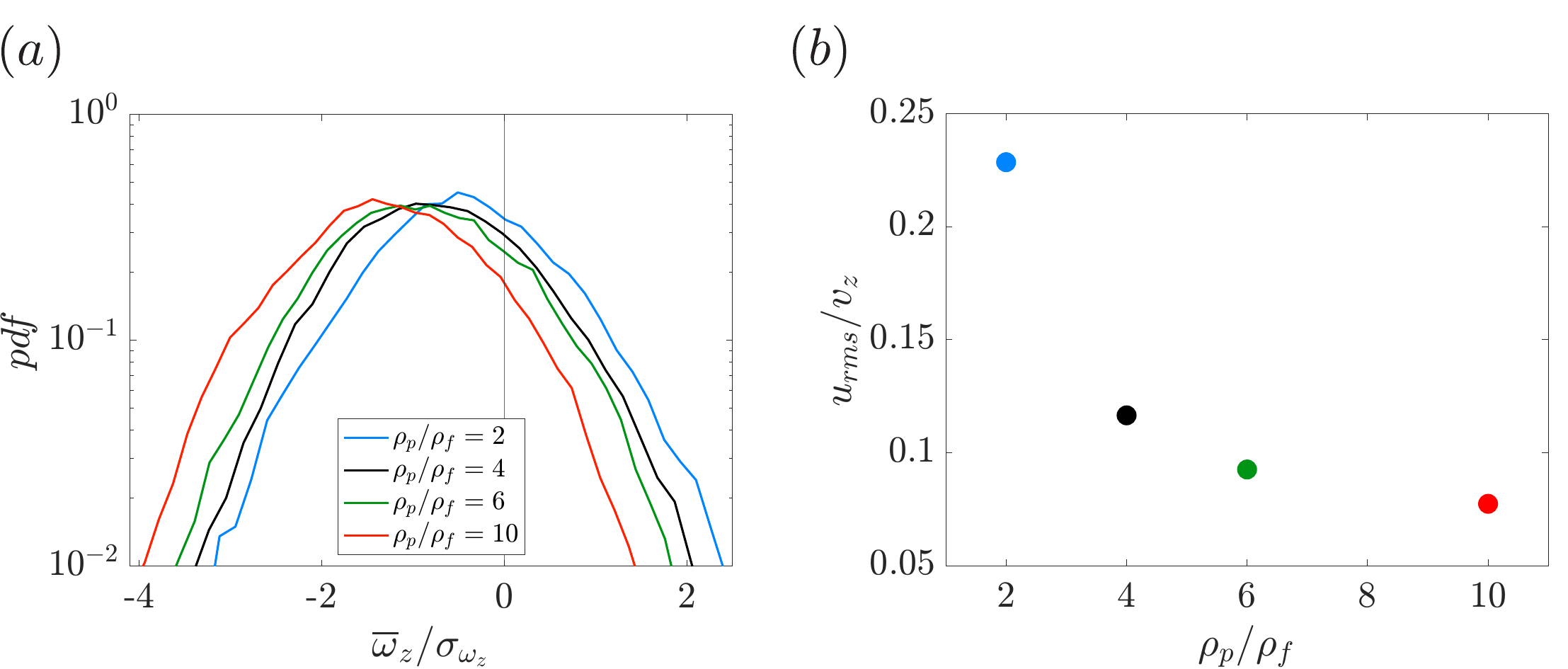}
\caption{
Single chiral particle with $\rho_p/\rho_f$ from $2$ to $10$ 
($Fr$ from $1$ to $5/9$) in a turbulent flow
with the forcing of Re30 case of
 table \ref{tab:param}:
(a) Probability density function of the fluid vertical vorticity component.
(b) Ratio of turbulent fluctuation intensity and mean vertical particle velocity
as function of $\rho_p/\rho_f$. Note the very different vertical scale in figures \ref{fig:vordrv30}b and \ref{fig:vordrv}b.
}
\label{fig:vordrv30}
\end{figure}

We have repeated the same simulations as before, increasing the turbulence
forcing as for the case Re30 of table \ref{tab:param}. For the sake of conciseness,
we report only the results of figure \ref{fig:vordrv30}, which shows an effect of
stronger turbulence only on the vorticity histograms of the lighter particles. Accordingly 
the ratio  $u_{rms}/ v_z$   shows a substantial increase for 
$\rho_p/\rho_f=2$ and $4$ ($Fr=1$ and $2/3$) while it is essentially unchanged for the higher 
density ratios.

The picture emerging from the  simulations  
with a single chiral particle 
is that its fall injects additional energy in the fluid whose
turbulence is altered; in turn, particle falling and rotational velocities are
affected by the strength of velocity fluctuations. Therefore the system
dynamics emerges out of the balance of a complex two--way coupling between solid and fluid phases.
Nevertheless, a clear trend is observed such that stronger velocity fluctuations reduce the
effects of particle and its chirality on turbulence, while higher particle/fluid density ratios
increase them.

\subsection{Volume fraction effect}
\label{sec:phi}
\noindent In real systems, such as helically swimming active matter \citep{Liebchen22} or
plants aiming at maximising seeds transport and dispersal \citep{Mazzolai21}, multiple chiral
particles interact simultaneously among them and react back on the surrounding
fluid.
Accordingly, we have investigated how the system dynamics change when several
particles are placed in a turbulent environment and what is the effect of 
their volume fraction $\phi$.

The immediate effect of increasing the number of particles is to enhance
the forcing on the turbulence, owing to growing
energy injection in the fluid phase. However, the presence of multiple
particles 
increases also energy dissipation, by friction with solid surfaces and the small scale 
of their wakes. Therefore turbulence can be strengthened or suppressed, depending on
which of the two mechanisms prevails. 

Furthermore, the presence of obstacles in the bulk flow reduces the size attainable by the 
 largest flow scales resulting in further modifications of the energy spectrum.
Finally, the collision among complex shape particles 
favours prolongated interactions rather than impulsive rebounds, as in the case of spheres 
or other compact objects, and this changes the solid body dynamics and the interaction with the flow.

\begin{table}
\begin{tabular*}{\linewidth}{@{\extracolsep{\fill}} cccccccc} 
$N_p$ &  $\phi\ (\%)$ &  $\langle v_z\rangle$ & $\langle \dot \theta_y \rangle$ & $Re_\lambda$  & $u_{rms}$ & $\varepsilon$             &  $\overline{ \omega}_z/\sigma_{\omega_z}$   \\ \hline \hline
      $0$  &  $0$        &    --    &          --        & $33.9$        &    $0.27$   &  $7.5\times10^{-3}$    &     $ 0.00$                    \\
      $1$  &  $0.05$     &  $-1.18$ &        $-0.17$     & $31.9$        &    $0.27$   &  $8.0\times10^{-3}$    &     $-0.43$                    \\
      $5$  &  $0.25$     &  $-1.17$ &        $-0.17$     & $29.5$        &    $0.29$   &  $1.2\times10^{-2}$    &     $-0.86$                    \\
      $10$ &  $0.50$     &  $-1.14$ &        $-0.17$     & $28.1$        &    $0.30$   &  $1.6\times10^{-2}$    &     $-1.21$                    \\
      $20$ &  $1.00$     &  $-1.10$ &        $-0.16$     & $24.4$        &    $0.31$   &  $2.5\times10^{-2}$    &     $-1.59$                    \\
      $40$ &  $2.00$     &  $-1.03$ &        $-0.16$     & $22.9$        &    $0.34$   &  $4.0\times10^{-2}$    &     $-1.67$                    \\
\end{tabular*}
\caption{Flow parameters for multiple particle simulations with variable volume fraction $\phi$,
 $\rho_p/\rho_f=2$ ($Fr=1$) and turbulent forcing as case Re30 of  table \ref{tab:param}. Where $\langle ~ \rangle$ indicates an average over the particles.}
\label{tab:phi30}
\end{table}

\begin{figure}
\centering
\includegraphics[width=\textwidth]{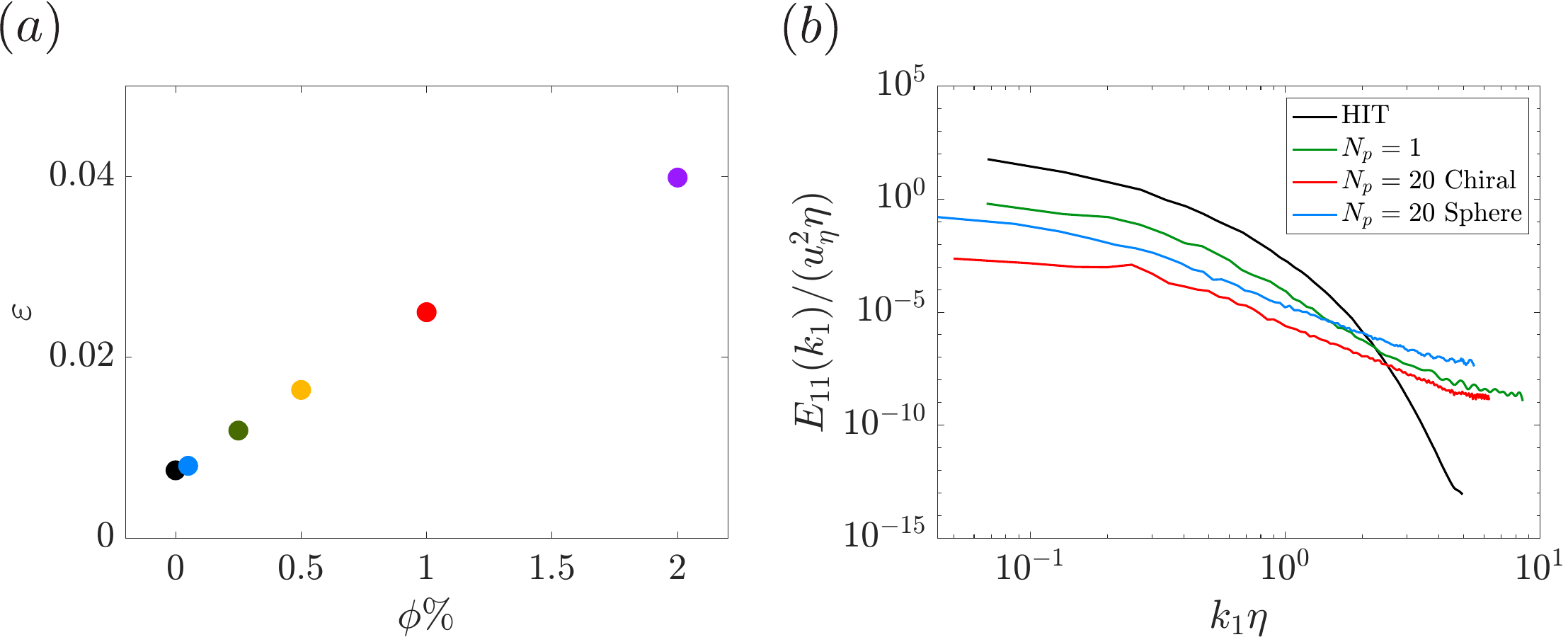}
\caption{
Multiple chiral particles  with $\phi$ from $0.05\%$ to $2\%$ in a turbulent flow
with the forcing of Re30 case of
 table \ref{tab:param} and $\rho_p/\rho_f=2$:
(a) Kinetic energy dissipation rate as function of the volume fraction $\phi$.
(b) Comparison of one--dimensional energy spectra for single--phase, one chiral particle,
$20$ chiral particles and $20$ spheres.
}
\label{fig:spephi30}
\end{figure}

Table \ref{tab:phi30} reports some representative data for flows with a turbulent forcing as in case Re30 of  table \ref{tab:param}
and an increasing number of particles with $\rho_p/\rho_f=2$ ($Fr=1$). It is indeed confirmed that a larger number of particles evolving 
within the same domain enhances velocity fluctuations, consistently with the augmented energy injection; however,  also viscous dissipation 
increases and this happens at a faster rate than the increase of $u_{rms}$.
The dependence of $\varepsilon$ on $\phi$ is displayed in figure \ref{fig:spephi30}a, which shows a linear behaviour which matches the findings of
\cite{Fornari19} for finite size spheres in turbulence.
The different growths of $\varepsilon$ and $u_{rms}$ with $\phi$ result in a decreasing 
$Re_\lambda$ indicating that, for this setup, more particles weaken turbulence by transferring more energy to small scales. 
This mechanism is quantitatively confirmed by the spectra of figure \ref{fig:spephi30}b, which shows an increased energy content 
at the smallest scales
and a reduced content at larger scales, which is similar to what was observed by \cite{Gao13} in HIT flows with spheres of 
diameter $\approx \lambda$.
The spectra of figure \ref{fig:spephi30}b also show that, for given $\phi$, the present chiral particles are much more 
effective than spheres in modifying turbulence. This might be due to the angular momentum of the dispersed
phase which in turn induces a net rotation of the fluid turbulence. Indeed, \cite{Biferale13} have already shown that,
in HIT, altering the nonlinear terms so to have a sign--definite helicity content heavily affects the energy cascade and
turbulence spectra.

A further consideration about the spectra of figure \ref{fig:spephi30}b is that, at a first sight, it seems possible to collapse the spectra for spheres and chiral particles by a simple rescaling although several direct tests have shown that this is not the case. In fact, we would like to stress that, chiral particles and spheres are equivalent only volumwise while the wetted surface of the former is bigger by $50\%$ and the front area by $25\%$. This implies that  features like energy dissipation by friction and the surrounding fluid dragged during the motion cannot be accounted for by a single scaling factor.  

\begin{figure}
\centering
\includegraphics[width=1.0\textwidth]{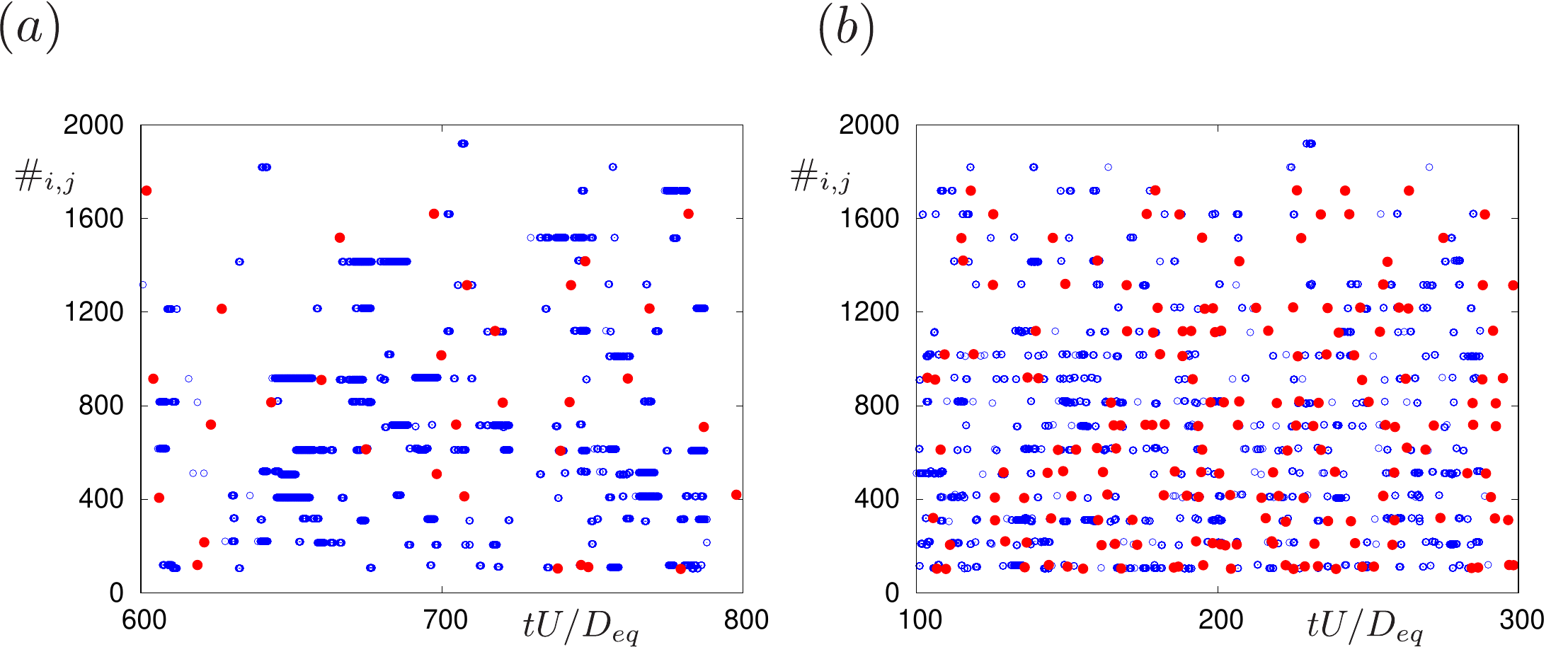}
\caption{
Collision events versus time for $N_p=20$ particles ($\phi=1\%$) in a turbulent flow:
a) forcing as Re30 case of table \ref{tab:param} and $\rho_p/\rho_f=2$;
b) forcing as Re60 case of table \ref{tab:param} and $\rho_p/\rho_f=7$;
the four--digit number on the y--axis identifies the two interacting particles:
for example, $\#i,j=0412$ indicates that particles $\#i=4$ and $\#j=12$ have collided.
Red bullets for spherical particles, blue open circles for chiral particles.
}
\label{fig:col30n}
\end{figure}

An additional reason for the marked difference between spheres and chiral particles could  be   the  collision 
dynamics which is illustrated in figure \ref{fig:col30n}.
A set of $N_p$ particles yields $N_p(N_p-1)/2$ distinct couples, if pairs $\#i$--$\#j$ and $\#j$--$\#i$
are equivalent. Whenever particles $\#i$ and $\#j$ are close enough to activate the collision
 model, we assemble a four--digit integer $\#ij$ from particles  identification number, 
and  record the time of interaction.
Figure \ref{fig:col30n} shows, in a representative time window, the interactions for $N_p=20$ spheres
and chiral particles with the same volume fraction $\phi=1\%$ for two flows with different turbulent forcing
 and density ratios $\rho_p/\rho_f$. It is evident that spheres 
tend to have short duration impulsive rebounds while chiral particles, with their complex shape,
favour entanglement which entail multiple contacts over several time units. Analysis of the flow fields
has evidenced an energy dissipation enhancement caused by particle collision which is consistent
with the energy spectra of figure \ref{fig:spephi30}b.

Clearly, frequency and duration of the collisions 
depend strongly on turbulence strength and particle density
(figure \ref{fig:col30n}). The comprehension of their statistics deserves a 
dedicated study which will be presented in a forthcoming paper.

\begin{figure}
\centering
\includegraphics[width=\textwidth]{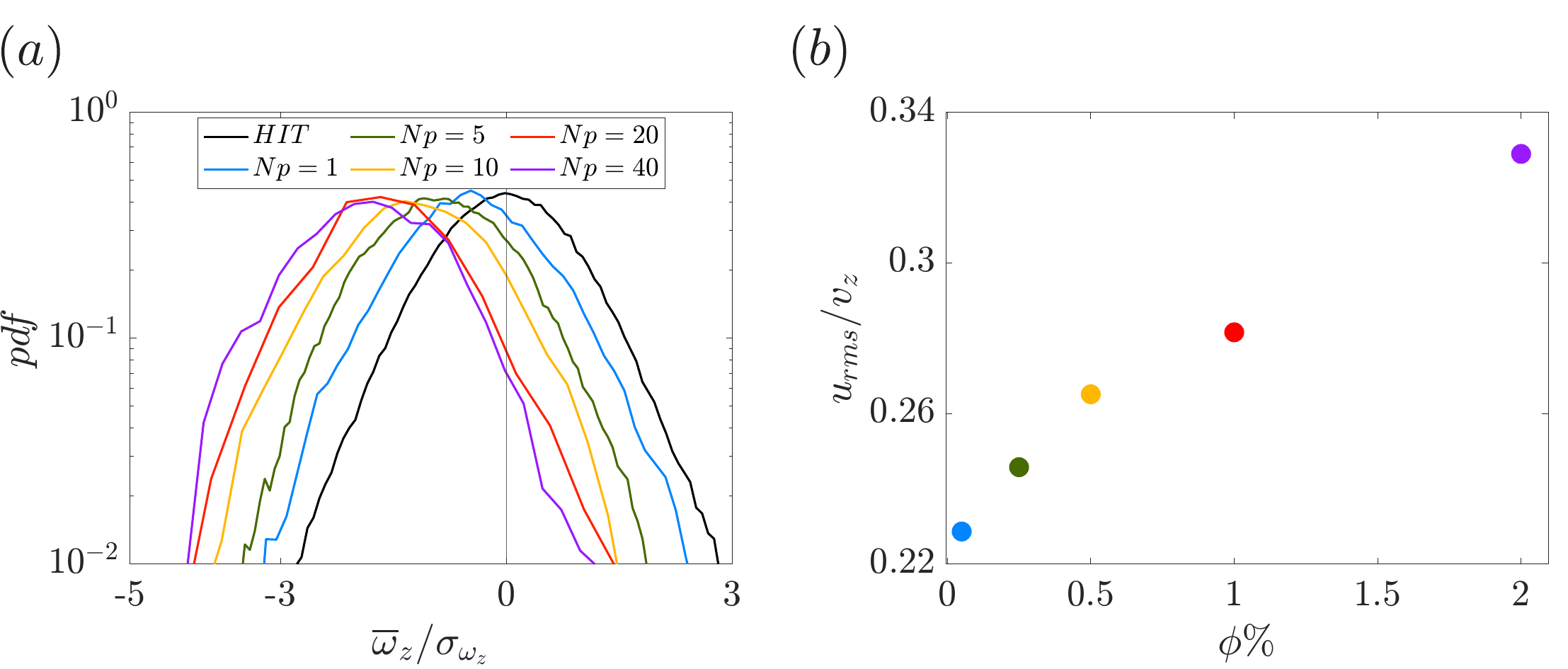}
\caption{
Multiple chiral particles with $\phi$ from $0.05\%$ to $2\%$ in a turbulent flow
with the forcing of Re30 case of
 table \ref{tab:param} and $\rho_p/\rho_f=2$:
(a) Probability density function of the fluid vertical vorticity component.
(b) Ratio of turbulent fluctuation intensity and mean vertical particle velocity
as function of the volume fraction $\phi$.
}
\label{fig:vorphi30}
\end{figure}

Concerning the particle angular velocity, table \ref{tab:phi30} shows that it 
is essentially constant with $\phi$ while the amplitude of
the mean fluid vorticity increases (figure \ref{fig:vorphi30}a); 
this can be explained by considering that $\langle \dot \theta_y \rangle$ is the mean
angular velocity of {\it each} particle. When their number increases the fluid must rotate faster to compensate for the total angular momentum
which yields a global zero circulation.

The values of $\langle v_z\rangle$ slightly decrease with $\phi$. This result agrees with the findings of
\cite{Fornari16} for spheres at similar volume fractions.
The ratio $u_{rms}/\langle |v_z|\rangle$ is however well below unity for all volume fractions 
(figure \ref{fig:vorphi30}b) and the vorticity histograms of figure
\ref{fig:vorphi30}a behave consistently, showing always asymmetric vorticity distributions.

After having investigated the effects of the single governing parameters separately,
we can now vary them simultaneously and verify whether the system behaves according to the
gained understanding. The picture emerging from the previous simulations is that particles
fall because of gravity and increase turbulent fluctuations, by
agitating the surrounding fluid at the expenses of their potential energy. At the same
time, also turbulent energy dissipation is enhanced. Thus the overall turbulence strength
($Re_\lambda$) results from the balance of these two competing effects. On the other hand,
particles settling velocity decreases as the turbulence strengthens, thus giving a two--way coupled 
dynamics. The chiral shape of the present particles couples rotational and translational
degrees of freedom which entails a net angular momentum of the solid phase and, correspondingly,
a mean vorticity of the fluid turbulence. Finally, increasing the volume fractions $\phi$ of the
chiral particles enhances the previous effects but introduces also a complex collision dynamics
which further complicates the interactions.

\begin{figure}
\centering
\includegraphics[width=\textwidth]{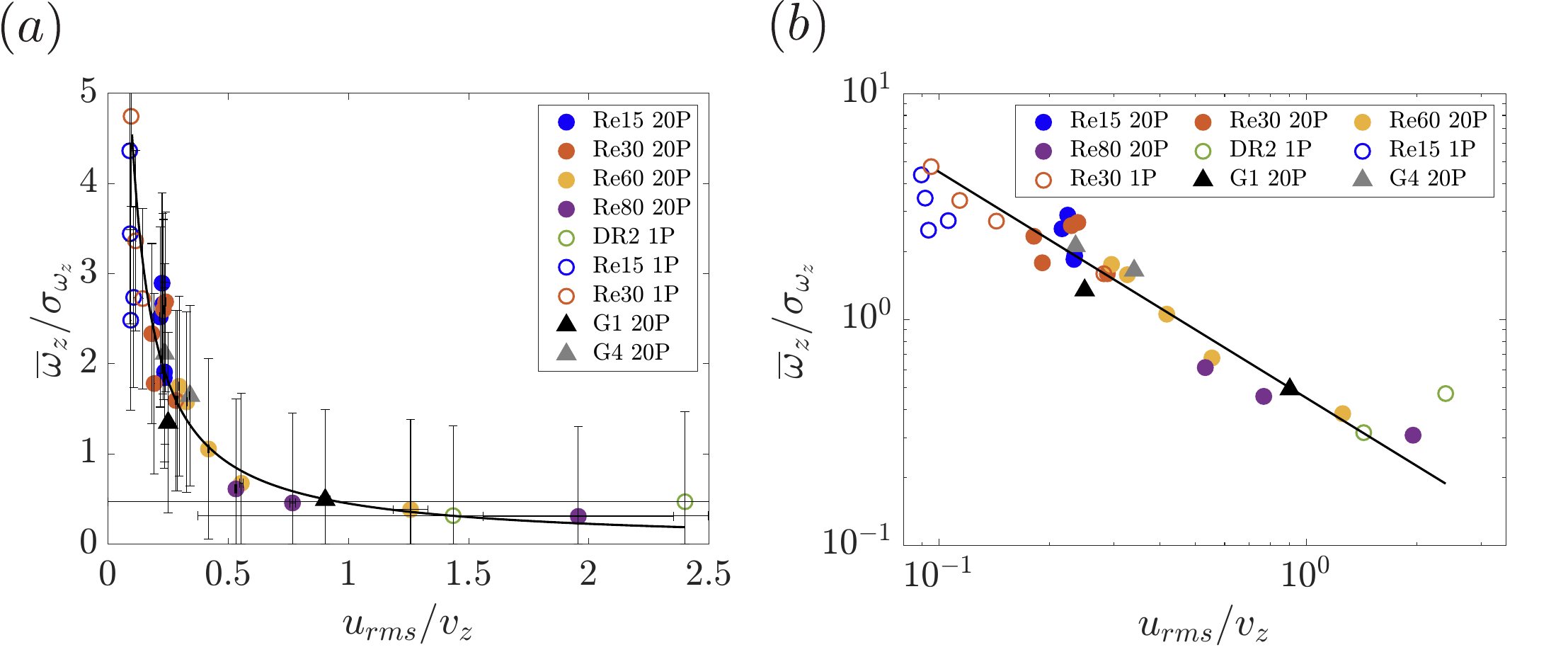}
\caption{
Mean flow vorticity normalised by its standard deviation versus ratio of turbulent velocity fluctuations 
with mean settling particle velocity in linear (a) and logarithmic scale (b). The black solid line is the 
curve $0.45 (u_{rms}/v_z)^{-1}$ which best fits all the simulations.}
\label{fig:allpa} 
\end{figure}

On account of the above arguments, the relative strength of particle forcing and turbulence
intensity can be quantified by the ratio $u_{rms}/\langle |v_z|\rangle$, while the chirality--induced 
flow rotation through the mean vorticity is normalised by its standard deviation 
$\langle \omega_z\rangle/\sigma_{\omega_z}$.
Several simulation campaigns for different values of $Re_\lambda$ and $\rho_p/\rho_f$ have
been performed. The results shown in figure \ref{fig:allpa} confirm that indeed the behaviour
is well described by these quantities for a wide range of governing parameters. Some additional
ad hoc cases have been run using different values of $\pmb{g}$ in order to vary independently
$Fr$ and $\rho_p/\rho_f$: also these data are reported in figure \ref{fig:allpa}. They
collapse with the rest of the simulations in which only $Fr$ has been varied as main parameter.

It is important to note that most of the flows of figure \ref{fig:allpa} have been computed with
$N_p=20$ particles ($\phi=1\%$) although we have verified that the same plot for all cases
obtained using $N_p=1$ ($\phi=0.05\%$) behaves similarly even if data collapse onto a different
curve. On the other hand, table \ref{tab:phi30} and the related discussion have evidenced that $\phi$ affects
the flow features in several ways and parametrising its mechanisms is not trivial. However, we can attempt
to estimate the effect of $\phi$ from table \ref{tab:phi30}, which has been obtained for a single value of
$Re_\lambda$ and $Fr$, and verify whether the same correction works for all other flows.
More in detail, from table \ref{tab:phi30} the quantities $u_{rms}/\langle |v_z|\rangle$ and $\langle \omega_z\rangle/\sigma_{\omega_z}$
at $N_p=1$ and $N_p=20$ have been employed to compute their ratios to be used as correction coefficients
for all cases with $N_p=1$ computed for different values of $Re_\lambda$ and $Fr$. 
Indeed, the open symbols in figure \ref{fig:allpa} confirm that the behaviour is the same and all
data now collapse on the same curve.

A final comment is in order concerning the huge error bar of the rightmost points in figure \ref{fig:allpa}a:
in this case the strength of turbulence (produced by the forcing  $\pmb{f}_{T}$ alone) is such to make
negligible the effect of the particle on the flow whose dynamics tend to that of a passive tracer. 
Accordingly, the settling velocity and the induced flow rotation should vanish while turbulent
fluctuations strengthen. We have found that achieving statistical convergence for these cases requires
very long simulation times (see also the large variations of the ratio $T_o/T_e$ reported in section
\ref{ssec:setup}) and for $u_{rms}/\langle |v_z|\rangle > 2.5$ obtaining convergence becomes unpractical. We have however 
verified that, regardless of turbulence intensity, simulations with $\rho_p/\rho_f=1$ always
yield $\langle |v_z|\rangle \rightarrow 0$ and $\langle \omega_z\rangle \rightarrow 0$, thus confirming that the data of figure 
\ref{fig:allpa} approach the correct limit.

\section{Conclusions and outlook}{\label{sec:conclusions}}
\noindent 
Using direct numerical simulations of homogeneous isotropic turbulence with finite--size, heavy chiral
particles, we have investigated the dynamics of the system under the influence of the main 
governing parameters which are: turbulence strength, solid--to--fluid density ratio (or Froude number) 
and volume fraction of the solid phase.

A particle, falling under the effect of an external gravity field, converts its potential energy into a
kinetic counterpart which is transferred to the fluid through viscous and pressure forces. On the one hand this
mechanism enhances the intensity of velocity fluctuations but, on the other hand, it also increases
the energy dissipation through viscous friction and the generation of small scales in the wake.
Depending on the specific flow parameters, the former or the latter mechanisms can prevail, thus boosting 
or suppressing turbulence intensity, respectively.

Turbulence of increasing strength, in turn, affects particle dynamics by reducing its mean falling velocity 
and decreasing the relevance of the additional energy input with respect to that introduced by the HIT
forcing alone.  

Particle chirality, by coupling translational and rotational degrees of freedom, produces a mean
flow vorticity which is evidenced through its probability density function with a non--zero peak
position.

Our results suggest that the ratio of turbulent velocity fluctuations to the mean particle settling velocity 
$u_{rms}/v_z$ can be used to characterise the intensity of the chirality induced mean rotation of the 
flow which decreases as the previous ratio increases: the implication of this finding is that particle
chirality does not matter much in   large-scale features of strong turbulence. 

The influence of the particle volume fraction $\phi$ is more complex than the other parameters since, 
in addition to the above mentioned
mechanisms, there are the interactions among particles and their collisions. The latter have shown to be
characterised by long entanglement times in contrast to spheres which instead rebound impulsively. 
These differences are evident also from the turbulent energy spectra
which, for a given volume fraction ($\phi=1\%$), show a more pronounced energy reduction caused
by chiral particles than by spheres.
Despite the more complex dynamics, yet the quantity $u_{rms}/v_z$ describes correctly the induced flow rotation
as confirmed by figure \ref{fig:allpa}.

This paper is only a first attempt to investigate the interaction of relatively complex shape chiral particles
with homogeneous isotropic turbulence and many things could be done to extend the study in different
ways:  more detailed flow statistics should be computed in order to understand
how turbulence is modified not only with respect to rotation.   Although the work done by the falling particles may be negligible for altering the energetics of larger scales in the flow, it could be significant at smaller scales. At these scales where the energy injection due to the particles is significant, the particle chirality might also matter. 
An extreme case confirming this observation is that for  $50\%$ mixtures of left- and right-handed particles, flow vorticity does not change although the small scale features of turbulence is altered.  

Heavier and lighter spheres in HIT (such as droplets in gases and droplets in liquids, respectively),
owing to centrifugal effects, are known to collect in different flow regions. Thus chiral particles
lighter than the surrounding fluid might behave differently from the present ones.
In this paper we have verified that left-- and right--handed particles behave identically except for
the sign of the induced flow rotation. A $50\%$ mixture of left-- and right--handed particles is expected
to produce zero net rotation of the flow although it is likely to alter other turbulence properties
very differently from a set of spheres with the same $\phi$.
Finally, the statistics of particle collisions have shown to strongly depend on the various flow parameters
(figure \ref{fig:col30n}). A deeper understanding of their features would be very useful to characterise
the dynamics of real suspensions.

All these points are currently under investigation or will be dealt with in the near future. They
will be the subject of forthcoming papers.

\backsection[Acknowledgements]{We wish to thank Mr. Xander M. de Wit and Prof. Federico Toschi for many 
fruitful clarifying discussions about the dynamics of chiral bodies and their interaction with turbulence.}

\backsection[Funding]{This research has received funding from the Dutch Research Council under the project ``Shaping turbulence with smart particles'' (file number OCENW.GROOT.2019.03). We acknowledge the access to several computational resources, all of which were used for this work: European High Performance Computing Joint Undertaking for awarding us access to Discoverer under the project EHPC-REG-2022R03-208 and the national e-infrastructure of SURFsara, a subsidiary of SURF cooperation.}

\backsection[Declaration of interests]{The authors report no conflict of interest.}

\backsection[Data availability statement]{The data that support the findings of this study are available upon reasonable request.}


\appendix
\section{Density ratio independence if $Fr=1$}\label{sec:appA}
\begin{figure}
\centering
\includegraphics[width=1.0\textwidth]{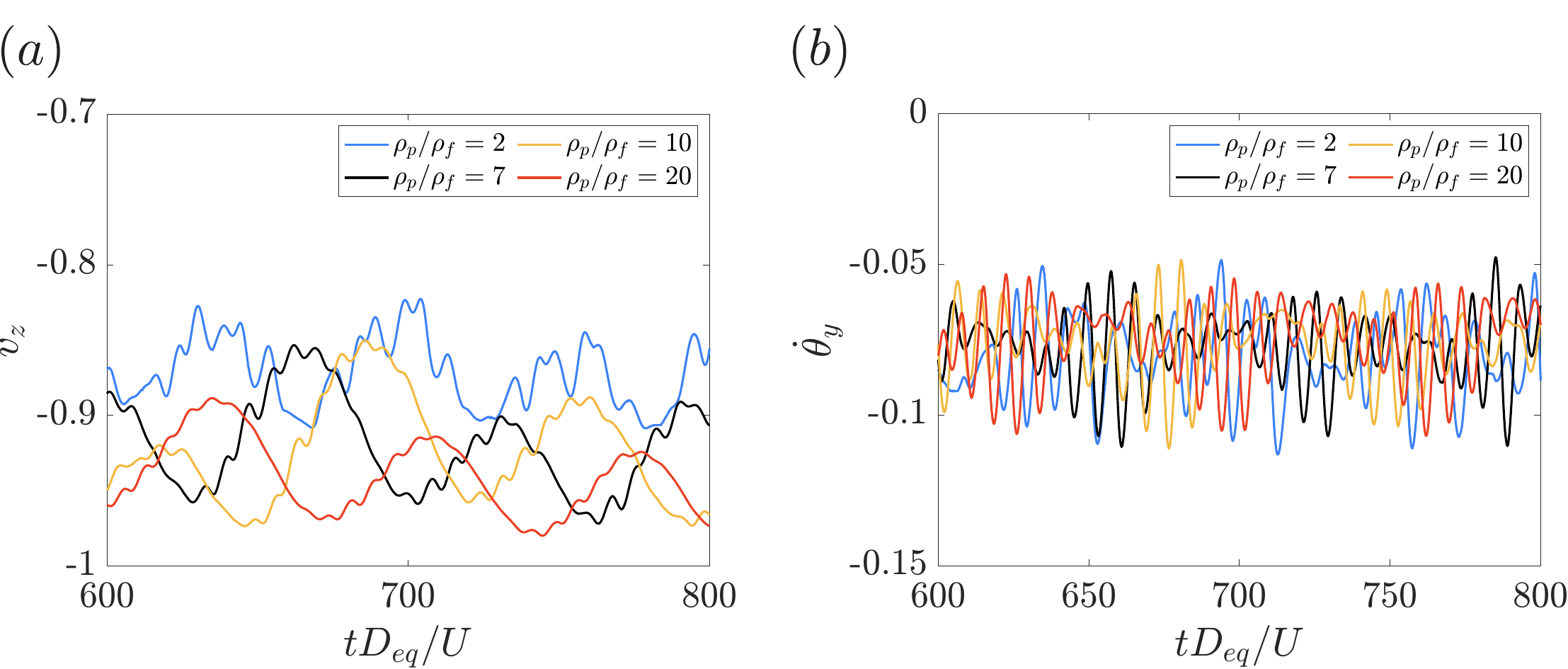}
\caption{
Single chiral particle falling in a stagnant fluid at $Re_p\approx 100$ for different $\rho_p/\rho_f$ and fixed $Fr=1$.
(a) Time evolution of the centre of mass vertical velocity component. 
(b) Time evolution of angular velocity $y-$component expressed 
in the particle body frame (green axes in figure \ref{fig:config}).}
\label{fig:appA1}
\end{figure}

In this appendix, we provide evidence of the reduced sensitivity of equations (\ref{eq:newton2}) 
to density ratio compared to the Froude number. Figure \ref{fig:appA1} reports some cases run for 
$2 \leq \rho_p/\rho_f \leq 20$ and $Fr=1$, showing only limited  differences. Indeed, some effects are
evident for the settling velocity, especially for the lowest density ratio. These however have to
be compared with the results of figure  \ref{fig:drv} where a smaller variation of $Fr$ produces much
larger effects.

As an aside we note that if  $U_g=\sqrt{D(\rho_f/\rho_p-1)|\pmb{g}|}$   is used as scaling velocity 
for all equations (rather than the flow velocity $U$) the Froude number turns out to be always $Fr=1$
and the resulting system show only limited sensitivity to the density ratio $\rho_p/\rho_f$.

\section{Kinetic energy balance}\label{appB}
\begin{figure}
\centering
\includegraphics[width=\textwidth]{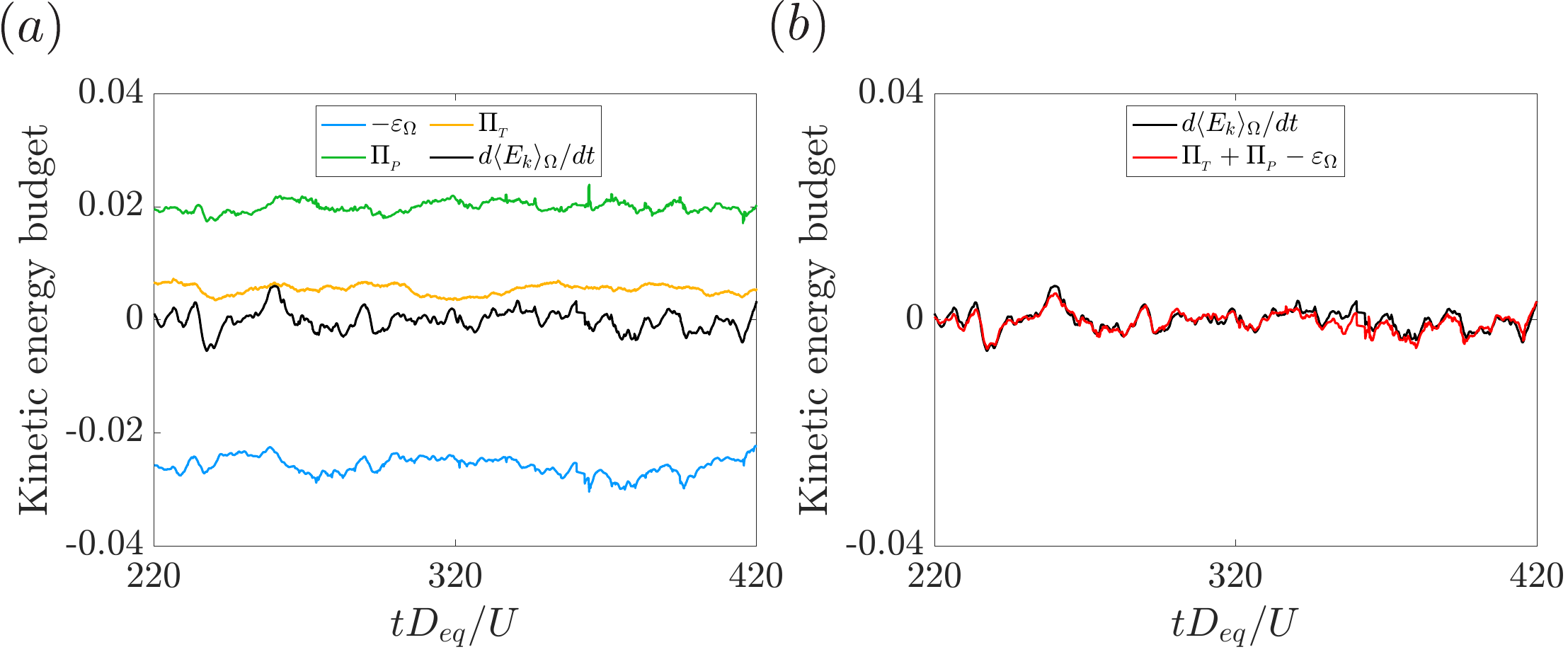}
\caption{Time evolution of left-- and right--hand side of equation (\ref{eq:kbal})  for $N_p=20$ chiral particles 
of $\rho_p/\rho_f=2$ in the forcing case Re30 of table \ref{tab:param}.{The right--hand side terms are shown both individually (a) and in sum (b).}}
\label{fig:kbal}
\end{figure}

\noindent As a further consistency check of the numerics,  we consider the energy budget obtained from equations
 (\ref{eq:NS}) after a scalar product with $\bold{u}$ and integrating over the entire domain $\Omega$.
We obtain, for the kinetic energy $E_k=\bold{u\cdot u}/2$, the following relation:
\begin{equation}
\label{eq:kbal}
\frac{d}{dt}\langle E_k\rangle_{_\Omega}=-\varepsilon_\Omega + \Pi_{_T}+\Pi_{_P}
\end{equation}
with the dissipation rate
\begin{equation}
\varepsilon_\Omega=\frac{1}{Re}\Biggl \langle  \Bigg|\frac{\partial \bold{u}}{\partial \bold{x}}  \Bigg|^2 \Biggr \rangle_\Omega,
\end{equation}
the work done by the turbulence forcing $\Pi_{_T}$ 
\begin{equation}
\Pi_{_T}=\langle \bold{u\cdot} \pmb{f}_{_T}\rangle_{_\Omega},
\end{equation}
and the fluid--particle coupling term $\Pi_{_P}$,
\begin{equation}
\Pi_{_P}=\langle \bold{u\cdot} \pmb{f}_{_P}\rangle_{_\Omega}.
\end{equation}
figure\ref{fig:kbal} shows the time evolution of the left-- and right--hand side of equation (\ref{eq:kbal}),
 confirming the balance.

\section{Computational domain size in gravity direction}\label{appC}
\begin{figure}
\centering
\includegraphics[width=1.0\textwidth]{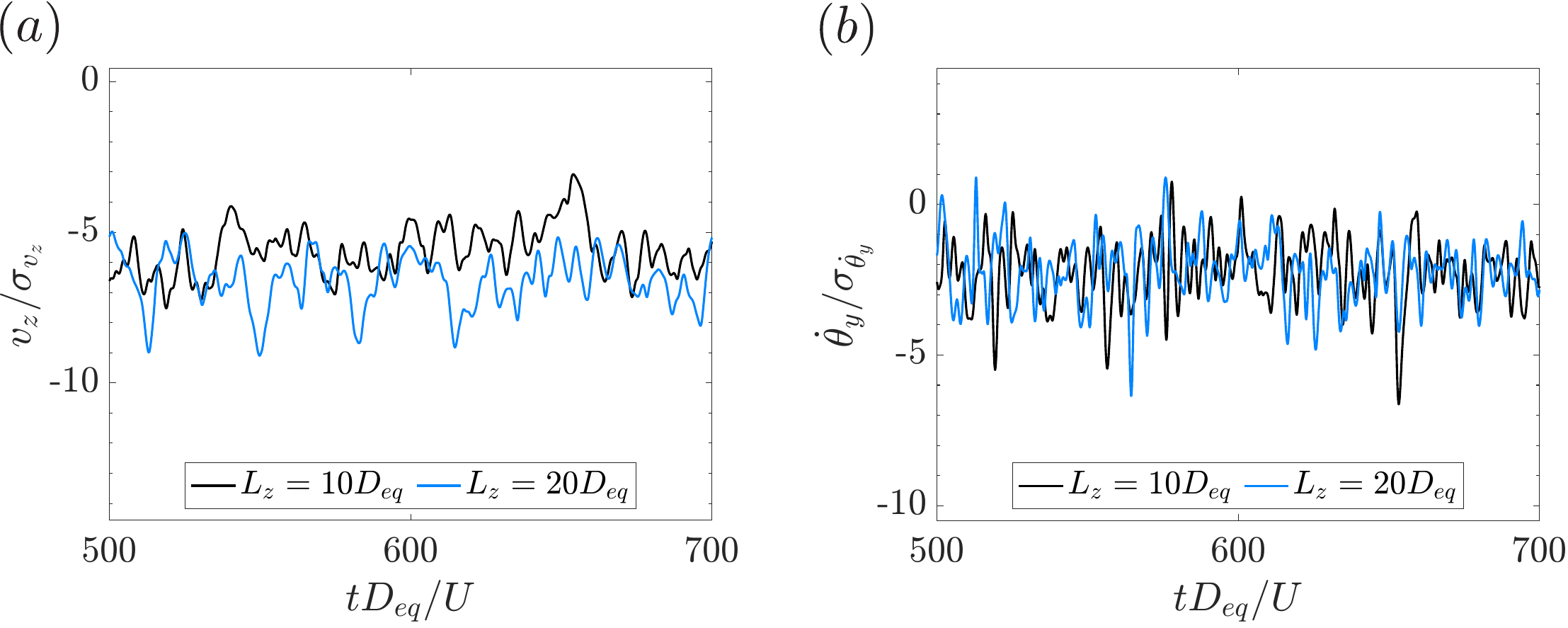}
\caption{
Single chiral particle of $\rho_p/\rho_f=2$ in a turbulent flow with the forcing of Re30 case of table\ref{tab:param}:
(a) Time evolution of the centre of mass vertical velocity component. 
(b) Time evolution of angular velocity $y-$component expressed 
in the particle body frame (green axes in figure \ref{fig:config}).}
\label{fig:appC2}
\end{figure}

\noindent Here we provide the results of additional simulations run in a computational domain whose size
has been doubled in the gravity direction ($L_z=20D_{eq}$). These tests were run to ascertain a possible 
influence of the vertical domain size on account of particles settling velocity. Figure \ref{fig:appC2} 
shows the time evolution settling and angular velocity for a single particle at $\rho_p/\rho_f=2$ in a 
turbulent flow with the forcing as case Re30 of table \ref{tab:param}: their mean values are 
$v_z/\sigma_{v_z} = -6.7 \pm 1.0$,  $-2.3 \pm 0.3$
for $L_z=20D_{eq}$ and $v_z/\sigma_{v_z} = -5.8 \pm 1.1$,  $-2.1 \pm 0.3$ for $L_z=10D_{eq}$, thus 
confirming that the cubic domain is indeed
adequate for our analyses.

\begin{figure}
\centering
\includegraphics[width=0.9\textwidth]{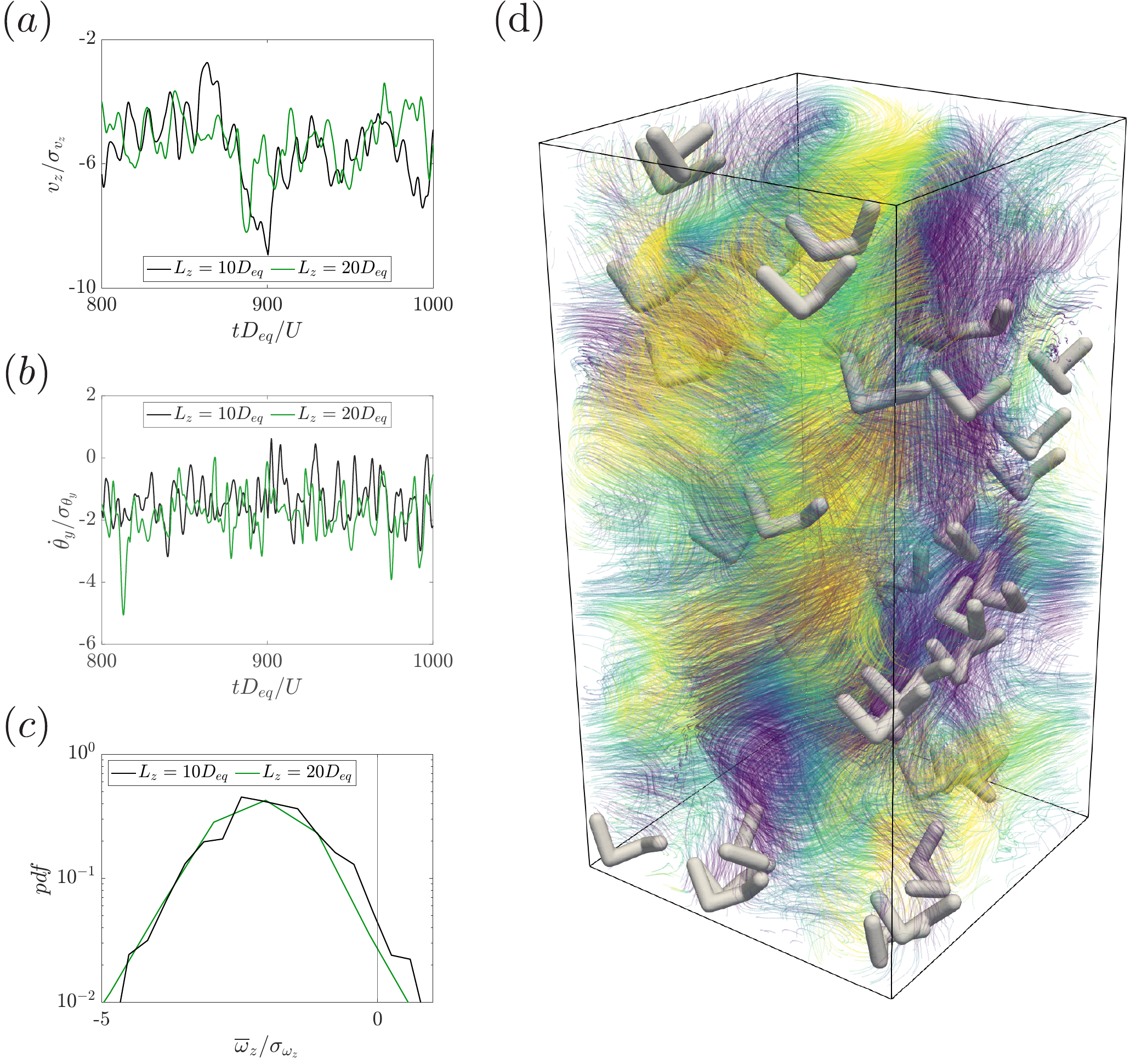}
\caption{
Crowd of chiral particles  of $\rho_p/\rho_f=2$ in a turbulent flow with the forcing of Re30 case of table\ref{tab:param}:
(a) Time evolution of the centre of mass vertical velocity component. 
(b) Time evolution of angular velocity $y-$component expressed 
in the particle body frame (green lines in figure \ref{fig:config}), 
(c) Probability density function of the fluid vertical vorticity component, (d) {streamlines of the nondimensional velocity field coloured according to the vertical component ranging from $-0.3$ (blue) to $0.3$ (yellow).}}
\label{fig:appC}
\end{figure}

In
figure \ref{fig:appC} we report the results of a similar test performed for a crowd of particles at 
$\rho_p/\rho_f=2$ in a turbulent flow with the forcing as case Re30 of table \ref{tab:param} and 
$\phi=1\%$ ($N_p=40$); once again the results show very good agreement with the analogous ones obtained
in a cubic computational domain.


\clearpage
\bibliographystyle{jfm}
\bibliography{refs}

\end{document}